\documentclass{aa}
\usepackage{graphicx}
\usepackage{txfonts}
\usepackage{natbib}
\bibpunct{(}{)}{;}{a}{}{,} 

\usepackage[figuresright]{rotating}
\usepackage{subfigure}
\newcommand{\kms}{km s$^{-1}$}
\newcommand{\ps}{s$^{-1}$}
\newcommand{\Q}{molecules~s$^{-1}$}

\begin{document}
   \title{Spectral Analysis of the \emph{Chandra} Comet Survey}

   \author{D.~Bodewits\inst{1}
          \and
          D.~J.~Christian\inst{2}
          \and
          M.~Torney\inst{3}
          \and
          M.~Dryer\inst{4}
          \and
          C.~M.~Lisse\inst{5}
          \and
          K.~Dennerl\inst{6}
          \and
          T.~H.~Zurbuchen\inst{7}
          \and
          S.~J.~Wolk\inst{8}
          \and
          A.~G.~G.~M.~Tielens\inst{9}
          \and
          R.~Hoekstra\inst{1}
          }

   \offprints{D. Bodewits}

   \institute{\textsc{kvi} atomic physics, University of Groningen, Zernikelaan 25, NL-9747 AA Groningen, The Netherlands\\
              \email{bodewits@kvi.nl, hoekstra@kvi.nl}
        \and
            Queen's University Belfast, Department of Physics and Astronomy, Belfast, BT7 1NN, UK\\
            \email{d.christian@qub.ac.uk}
         \and
            Atoms Beams and Plasma Group, University of Strathclyde, Glasgow, G4 0NG, UK\\
            \email{torney@phys.strath.ac.uk}
         \and
            \textsc{noaa} Space Environment Center, 325 Broadway, Boulder,
            CO 80305, USA\\
            \email{murray.dryer@noaa.gov}
         \and
            Planetary Exploration Group, Space Department, Johns Hopkins University Applied Physics Laboratory, 11100 Johns Hopkins Rd, Laurel, MD 20723, USA\\
            \email{carey.lisse@jhuapl.edu}
        \and
            Max-Planck-Institut f\"{u}r extraterrestrische Physik, Giessenbachstrasse, 85748 Garching, Germany\\
            \email{kod@mpe.mpg.de}
         \and
            The University of Michigan, Department of Atmospheric, Oceanic and Space Sciences, Space Research Building, Ann Arbor, MI 48109-2143, USA\\
            \email{thomasz@umich.edu}
         \and
            Harvard-Smithsonian Center for Astrophysics, 60 Garden Street, Cambridge, MA 02138, USA \\
            \email{swolk@head.cfa.harvard.edu}
         \and
            \textsc{nasa} Ames Research Center, MS 245-3, Moffett Field, CA 9435-1000, USA \\
            \email{tielens@astro.rug.nl}
  }

   \date{Received \today}

\abstract {} {We present results of the analysis of cometary X-ray
spectra with an extended version of our charge exchange emission
model (Bodewits et al. 2006). We have applied this model to the
sample of 8 comets thus far observed with the \emph{Chandra} X-ray observatory
and \textsc{acis} spectrometer in the 300--1000~eV range. The surveyed comets
are C/1999~S4~(\textsc{linear}), C/1999 T1 (McNaught--Hartley), C/2000 WM1 (\textsc{linear}),
153P/2002 (Ikeya--Zhang), 2P/2003 (Encke), C/2001 Q4 (\textsc{neat}),
9P/2005 (Tempel~1) and 73P/2006-B (Schwassmann--Wachmann 3) and the
observations include a broad variety of comets, solar wind
environments and observational conditions.} {The interaction model is based on state selective, velocity dependent charge exchange cross sections and is used to explore how cometary X-ray emission depend on cometary,
observational and solar wind characteristics. It is further demonstrated that cometary X-ray spectra mainly reflect the state of the local solar wind. The current sample of \emph{Chandra} observations was fit using the constrains of the charge exchange model, and relative solar wind abundances were derived from the X-ray spectra. } {Our analysis
showed that spectral differences can be ascribed to different solar wind states, as such identifying comets interacting with (I) fast, cold wind,
(II), slow, warm wind and (III) disturbed, fast, hot winds associated with
interplanetary coronal mass ejections. We furthermore predict the
existence of a fourth spectral class, associated with the cool,
fast high latitude wind.} {}


   \maketitle

%
\section{Introduction}

When highly charged ions from the solar wind collide on a neutral
gas, the ions get partially neutralized by capturing electrons
into an excited state. These ions subsequently decay to the
ground state by the emission of one or more photons. This photon
emission is called charge exchange emission (\textsc{cxe}) and it has been
observed from comets, planets and the interstellar medium in X-rays and the
Far-UV \cite{Lis96, Kra97, Sno04, Den02}. The spectral shape of
the \textsc{cxe} depends on properties of both the neutral gas and the
solar wind and the subsequent emission can therefore be regarded
as a fingerprint of the underlying interactions \cite{Cra97,
Kha00, Kha01, Bei03, Bod04a, Bod06}.

Since the first observations of cometary X-ray emission, more than 20 comets have been observed with
various X-ray and Far-UV observatories \cite{Lis04, Kra04b}.
This observational sample contains a broad variety of comets,
solar wind environments and observational conditions. The
observations clearly demonstrate the diagnostics available from
cometary charge exchange emission.

First of all, the emission morphology is a tomography of the
distribution of neutral gas around the nucleus \cite{Weg04}.
Gaseous structures in the collisionally thin parts of the coma
brighten, such as the jets in 2P/Encke \cite{Lis05}, the Deep
Impact triggered plume in 9P/Tempel~1 \cite{Lis07} and the
unusual morphology of comet 6P/d'Arrest \cite{Mum97}. In other
comets, the X-ray emission clearly mapped a spherical gas
distribution. This resulted in a characteristic crescent shape for
larger and hence collisionally thick comets observed at phase
angles of roughly 90 degrees (e.g. Hyakutake - \citet{Lis96},
\textsc{linear} S4 - \citet{Lis01}). Macroscopic features of the plasma interaction
such as the bowshock are observable, too \cite{Weg05}.

Secondly, by observing the temporal behavior of the comet's
X-ray emission, the activity of the solar wind and comet can be
monitored. This was first shown for comet C/1996~B2~(Hyakutake)
\cite{Neu00} and recently in great detail by long term
observations of comet 9P/2005~(Tempel~1) \cite{Wil06, Lis07} and
73P/2006~(Schwassmann--Wachmann~3C) \cite{Bro07}, where cometary
X-ray flares could be assigned to either cometary outbursts
and/or solar wind enhancements.

Thirdly, cometary spectra reflect the physical
characteristics of the solar wind; e.g. spectra resulting from
either fast, cold (polar) wind and slow, warm equatorial solar
wind should be clearly different \cite{Sch00,Kha01,Bod04a}.
Several attempts were made to extract ionic abundances from the
X-ray spectra.

The first generation spectral models have all made
strong assumptions when modelling the X-ray spectra \cite{Hab97, Weg98, Kha00, Sch00,  Lis01, Kha01,Kra02, Bei03,
Weg04,Bod04a,Kra04,Lis05}. Here, we present a more elaborate and sophisticated procedure to
analyze cometary X-ray spectra based on atomic physics input,
which for the first time allows for a comparative study of all
existing cometary X-ray spectra. In Section \ref{sec:model},
our comet-wind interaction model is briefly introduced. In
Section \ref{sec:modelresults}, it is demonstrated how cometary
spectra are affected by the velocity and target dependencies of
charge exchange reactions. In Section \ref{sec:obs}, the various
existing observations performed with the \emph{Chandra} X-ray
Observatory, as well as the solar wind data available are
introduced. Based upon our modelling, we construct an analytical method of which the details and results are presented
in Section~\ref{sec:fits}. In Section \ref{sec:discussion}, we
discuss our results in terms of comet and solar wind
characteristics. Lastly, in Section~7 we summarize our findings. Details of the individual \emph{Chandra} comet observations are given in Appendix~A.


\section{Charge Exchange Model}\label{sec:model}

\subsection{Atomic structure of He-like ions}
   \begin{figure}
   \centering
    \includegraphics[width=8cm]{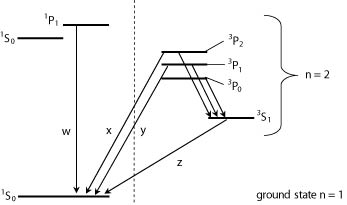}
   \caption{Part of the decay scheme of a helium--like ion. The $^1$S$_0$ decays to the ground state via two-photon processes (not indicated). \label{fig:grotrian}}
    \end{figure}
Electron capture by highly charged ions populates highly excited
states, which subsequently decay to the ground state. These
cascading pathways follow ionic branching ratio statistics.
Because decay schemes work as a funnel, the lowest transitions
($n=2 \to 1$) are the strongest emission lines in \textsc{cxe} spectra. For
helium-like ions, these are the forbidden line ($z$:
1s$^2$~$^1$S$_0$--1s2s $^3$S$_1$), the intercombination lines
($y,x$: 1s$^2$~$^1$S$_0$--1s2p $^3$P$_{1,2}$), and the resonance
line (w: 1s$^2$~$^1$S$_0$--1s2p $^1$P$_1$), see
Figure~\ref{fig:grotrian}.

The apparent branching ratio, B$_{\mathrm{eff}}$, for the intercombination transitions
is determined by weighting branching ratios (B$_j$) derived from theoretical transition rates compiled
by \citet{Por00,Por01}, by an assumed statistical population of
the triplet P-term: \begin{equation} B_{\mathrm{eff}} =
\sum^2_{j=0}\frac{(2j+1)}{(2L+1)(2S+1)} \cdot\mathrm{B}_j
\end{equation} The resulting effective branching ratios are given
in Table~\ref{tab:branch_helium}. These ratios can only be observed at conditions where the metastable state is not destroyed (e.g. by UV flux or collisions) before it decays. In contrast to many other astrophysical X-ray sources, this condition is fulfilled in cometary atmospheres, making the forbidden lines strong markers of \textsc{cxe} emission.

\begin{table}
\caption[]{Apparent effective branching ratios (B$\mathrm{_{eff}}$) for the
relaxation of the 2$^3$P-state of He-like carbon, nitrogen,
oxygen and neon. \label{tab:branch_helium}} \centering
\begin{tabular}{c c c c c}
\hline\hline
transition      &   \ion{C}{v}  &   \ion{N}{vi} &   \ion{O}{vii}  & \ion{Ne}{ix}  \\
\hline
1s$^2$ ($^1$S$_0$)--1s2p ($^3$P$_{1,2}$)    &   0.11    &   0.22    &   0.30  &  0.34\\
1s2s ($^3$S$_1$)--1s2p ($^3$P$_{0,1,2})$ &   0.89    &   0.78    &   0.70   & 0.66\\
\hline
\end{tabular}
\end{table}

\subsection{Emission Cross Sections}
To obtain line emission cross sections we start with an initial state population based on state selective electron capture cross sections and then track the relaxation pathways defined by the ion's branching ratios.

\begin{table*}
\caption[]{Compilation of theoretical, velocity dependent emission
cross sections for collisions between bare- and H-like solar wind
ions and atomic hydrogen, in units of 10$^{-16}$~cm$^2$. See text
for details. We estimate uncertainties to be ca.~20\%. The ion column contains the \emph{resulting} ion, not the original solar wind ion. Line energies compiled from \cite{Gar65,Vai85,Dra88,Sav03} and the \textsc{chianti} database \cite{Der97,Lan06}.
\label{tab:em_cross}} \centering
\begin{tabular}{c l l c c c c c}
\hline\hline $E$ (eV) & Ion  & Transition & 200 \kms & 400 \kms & 600 \kms & 800 \kms & 1000 \kms \\
\hline
299.0   &   \ion{C}{v}  &   z   &   8.7 &   12  &   16  &   18  &   20  \\
304.4   &   \ion{C}{v}  &   x,y &   0.65    &   1.0 &   1.5 &   1.7 &   1.8 \\
307.9   &   \ion{C}{v}  &   w   &   1.8 &   3.0 &   4.1 &   4.8 &   5.2 \\
354.5   &   \ion{C}{v}  &   1s3p-1s$^2$ &   0.55    &   0.71    &   0.81    &   1.0 &   1.3 \\
367.5   &   \ion{C}{v}  &   1s4p-1s$^2$ &   0.70    &   0.66    &   0.76    &   0.74    &   0.72    \\
367.5   &   \ion{C}{vi} &   2p-1s   &   15  &   26  &   30  &   33  &   34  \\
378.9   &   \ion{C}{v}  &   1s5p-1s$^2$ &   0.00    &   0.02    &   0.05    &   0.04    &   0.04    \\
419.8   &   \ion{N}{vi} &   z   &   13  &   23  &   28  &   29  &   29  \\
426.3   &   \ion{N}{vi} &   x,y &   2.7 &   4.3 &   5.3 &   5.7 &   6.0 \\
430.7   &   \ion{N}{vi} &   w   &   3.8 &   6.0 &   7.4 &   8.1 &   8.5 \\
435.5   &   \ion{C}{vi} &   3p-1s   &   1.6 &   4.0 &   4.7 &   4.7 &   4.8 \\
459.4   &   \ion{C}{vi} &   4p-1s   &   2.9 &   5.9 &   7.0 &   6.4 &   6.0 \\
471.4   &   \ion{C}{vi} &   5p-1s   &   0.55    &   1.0 &   1.3 &   0.85    &   0.54    \\
497.9   &   \ion{N}{vi} &   1s3p-1s$^2$ &   0.43    &   0.99    &   1.3 &   1.3 &   1.3 \\
500.3   &   \ion{N}{vii}    &   2p-1s   &   40  &   45  &   44  &   42  &   42  \\
523.0   &   \ion{N}{vi} &   1s4p-1s$^2$ &   0.81    &   1.6 &   1.9 &   1.8 &   1.7 \\
534.1   &   \ion{N}{vi} &   1s5p-1s$^2$ &   0.14    &   0.31    &   0.33    &   0.21    &   0.14    \\
561.1   &   \ion{O}{vii}    &   z   &   37  &   34  &   33  &   32  &   31  \\
568.6   &   \ion{O}{vii}    &   x,y &   10  &   10  &   10  &   9.9 &   9.7 \\
574.0   &   \ion{O}{vii}    &   w   &   9.9 &   11  &   11  &   11  &   10  \\
592.9   &   \ion{N}{vii}    &   3p-1s   &   6.3 &   4.9 &   4.8 &   4.5 &   4.3 \\
625.3   &   \ion{N}{vii}    &   4p-1s   &   2.9 &   2.9 &   3.7 &   4.3 &   4.6 \\
640.4   &   \ion{N}{vii}    &   5p-1s   &   11  &   5.2 &   3.7 &   2.7 &   2.2 \\
650.2   &   \ion{N}{vii}    &   6p-1s   &   0.00    &   0.21    &   0.13    &   0.09    &   0.08    \\
653.5   &   \ion{O}{viii}   &   2p-1s   &   27  &   40  &   48  &   51  &   53  \\
665.6   &   \ion{O}{vii}    &   1s3p-1s$^2$ &   1.7 &   1.3 &   1.3 &   1.2 &   1.2 \\
697.8   &   \ion{O}{vii}    &   1s4p-1s$^2$ &   0.81    &   0.79    &   1.0 &   1.2 &   1.3 \\
712.8   &   \ion{O}{vii}    &   1s5p-1s$^2$ &   2.8 &   1.3 &   0.92    &   0.68    &   0.54    \\
722.7   &   \ion{O}{vii}    &   1s6p-1s$^2$ &   0.00    &   0.06    &   0.04    &   0.02    &   0.02    \\
774.6   &   \ion{O}{viii}   &   3p-1s   &   2.6 &   4.7 &   5.6 &   5.3 &   5.0 \\
817.0   &   \ion{O}{viii}   &   4p-1s   &   1.0 &   1.6 &   2.0 &   2.2 &   2.3 \\
836.5   &   \ion{O}{viii}   &   5p-1s   &   2.4 &   4.0 &   4.6 &   4.1 &   3.7 \\
849.1   &   \ion{O}{viii}   &   6p-1s   &   1.6 &   1.6 &   1.5 &   1.1 &   0.67    \\

\hline
\end{tabular}
\end{table*}

Electron capture reactions can be strongly dependent on target
effects. An important difference between reactions with atomic
hydrogen and the other species is the presence of multiple
electrons, hence allowing for multiple (mostly double) electron
transfer. It has been demonstrated both experimentally and
theoretically that double electron capture can be an important
reaction channel in multi-electron targets and that after autoionization to an excited state it may contribute to the X-ray emission \cite{Ali05, Hoe89, Bei03,Otr06, Bod06}.
Unfortunately, experimental data on reactions with species typical
for cometary atmospheres, such as H$_2$O, atomic O and CO are at
best scarcely available. Because the first ionization potentials
of these species are all close to that of atomic H, using state
selective one electron capture cross sections for bare ions charge
exchanging with atomic hydrogen from theory is a reasonable
assumption, which is also confirmed by experimental studies
\cite{Gre00,Gre01,Bod06}. Here, we will use the working
hypothesis that effective one electron cross sections for multi-electron
targets present in cometary atmospheres are at least roughly comparable to cross
sections for one electron capture from H. Based on this
hypothesis, we will use our comet-wind interaction model to
evaluate the contribution of the different species.

For our calculations, we use a compilation of theoretical state
selective, velocity dependent cross sections for collisions with
atomic hydrogen \cite{Err04,Fri84,Gre82,Shi83}. We furthermore
assume that capture by H-like ions leads to a statistical triplet
to singlet ratio of 3:1, based on measurements by
\citet{Sur91,Bli98}. We will first focus on the strongest emission
features, which are the $n = 2 \to 1$~transitions, i.e., the
Ly-$\alpha$ transition (H-like ions) or the forbidden, resonance
and intercombination lines (He-like ions).

\begin{figure}
   \centering
    \includegraphics[width=8cm]{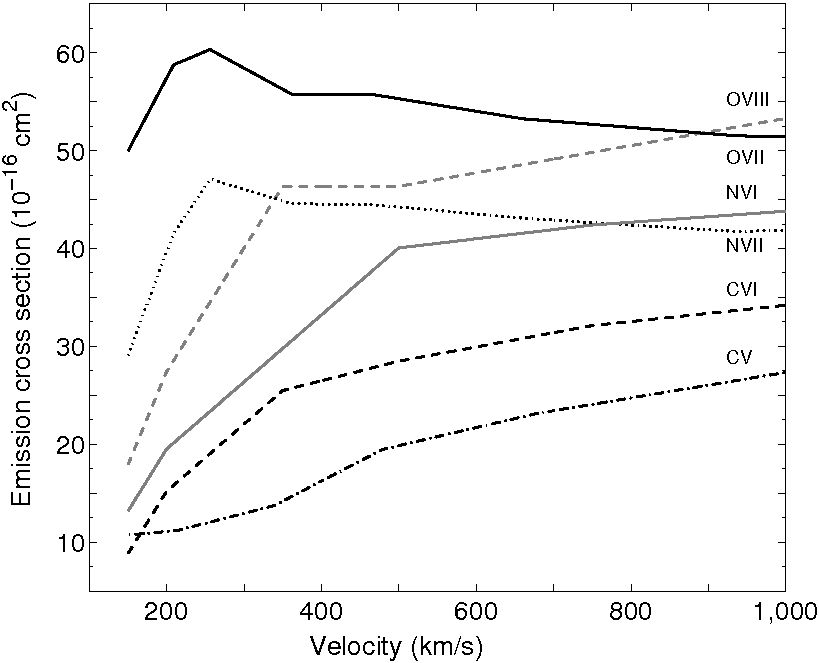}
   \caption{Velocity dependence of Ly-$\alpha$ or the sum of the forbidden/resonance/intercombination emission cross sections of different solar wind ions: \ion{O}{viii} (dashed, grey line), \ion{O}{vii} (solid, black line), \ion{N}{vii} (dotted, black line), \ion{N}{vi} (solid, grey line), \ion{C}{vi} (dashed, black line) and \ion{C}{v} (dash-dotted, black line).\label{fig:em_cross}}
    \end{figure}

In Fig.~\ref{fig:em_cross}, the emission cross sections of the
Ly-$\alpha$ or the sum of the emission cross sections of the forbidden, resonance and intercombination lines of
different ions (C, N, O) are shown as a function of collision
velocity, for one electron capture reactions with atomic hydrogen.
This figure sets the stage for solar wind velocity induced effects
in cometary X-ray spectra. Most important is the effect of the
velocity on the two carbon emission features; their prime emission
features increase by a factor of almost two when going from
typical `slow' to typical `fast' solar wind velocities. The
\ion{O}{viii} Ly-$\alpha$ emission cross section can be seen to
drop steeply below ca.~300~km~s$^{-1}$. The \ion{N}{vi}
K-$\alpha$ displays a similar, though somewhat less strong
behavior.

   \begin{figure}[t]
   \centering
    \includegraphics[width=8cm]{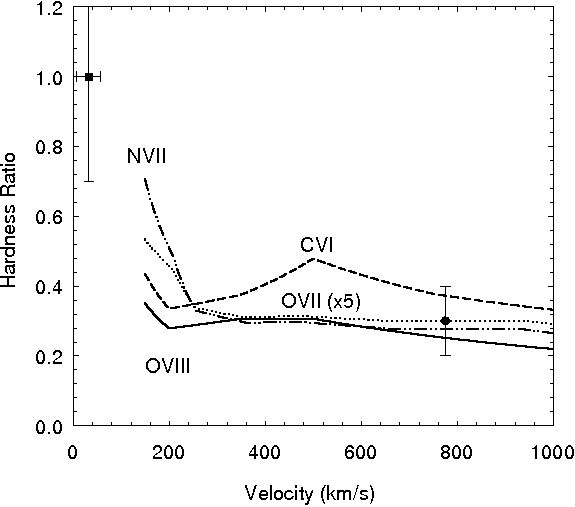}
   \caption{Velocity dependence of the hardness ratio of different solar wind ions: \ion{O}{viii} (solid line), \ion{O}{vii} (dashed line) \ion{N}{vii} (dashed line) and \ion{C}{vi} (dash-dotted line). Also shown are two experimentally obtained hardness ratios by \cite{Bei01} and \cite{Gre00} for O$^{8+}$ colliding on CO$_2$ and H$_2$O, respectively (see text).\label{fig:hardness}}
    \end{figure}

The relative intensity of the emission lines (per species) is governed by the state selective electron capture cross
sections of the charge exchange reaction and the branching ratios
of the resulting ion. A measure of these intensities is the
hardness ratio \citep{Bei01}, which is defined as the ratio between the emission
cross sections of the higher order terms of the Lyman-series and Ly-$\alpha$ (or between the higher order K-series and
K-$\alpha$ in case of He-like ions):

\begin{equation}\frac{\sum_{n>2}^\infty
\sigma_{\mathrm{em}}(\mathrm{Ly-}n)}{\sigma_{\mathrm{em}}(\mathrm{Ly-}\alpha)}
\end{equation}

For electron capture by H-like ions, we will use the ratio between
the sum of the resonance-, intercombination and forbidden emission
lines  and the rest of the K-series as the hardness ratio.
Fig.~\ref{fig:hardness} shows the hardness ratios of \textsc{cxe} from
abundant solar wind ions. The figure shows that most hardness
ratios are constant at typical solar wind velocities (above 300~\kms) but it also clearly demonstrates the suggestion made by
\citet{Bei01} that hardness ratios are good candidates for studies
of velocimetry deep within the coma when the solar wind has slowed
down by mass loading.

\subsection{Interaction Model}
Cometary high-energy emission depends upon certain properties of
both the comet (gas production rate, composition, distance to the
Sun) and the solar wind (speed, composition). Recently, we
developed a model that takes each of these effects into account
\cite{Bod06}, which we will briefly describe here.

The neutral gas model is based on the Haser-equation, which
assumes that a comet has a spherically
expanding neutral coma \cite{Has57, Fes81}. The lifetime of
neutrals in the solar radiation field varies greatly amongst
species typical for cometary atmospheres \cite{Hue92}. The
dissociation and ionization scale lengths also depend on absolute
UV fluxes, and therefore on the distance to the Sun. The coma
interacts with solar wind ions, penetrating from the sunward side
following straight line trajectories. The charge exchange
processes between solar wind ions and coma neutrals are explicitly
followed both in the change of the ionization state of the solar
wind ions and in the relaxation cascade of the excited ions (as
discussed above).

Due to its interaction with the cometary atmosphere, the solar wind
is both decelerated and heated in the bow shock. This bow shock
does not affect the ionic charge state distribution. The bow shock
lowers the drift velocity of the wind but at the same time
increases its temperature and the net collision velocity of the
ions is ca. 77$\%$ of the initial velocity v$(\infty)$ throughout
the interaction zone. We use a rule of thumb derived by
\citet{Weg04} to estimate the stand-off distance $R_{\mathrm{bs}}$ of the
bow shock.

Deep within the coma, the solar wind finally cools down as the hot
wind ions, neutralized by charge exchange, are replaced by cooler
cometary ions. For simplicity however, we shall assume that the
wind keeps a constant velocity and temperature after crossing the
bow shock.

   \begin{figure}
   \centering
    \includegraphics[width=8cm]{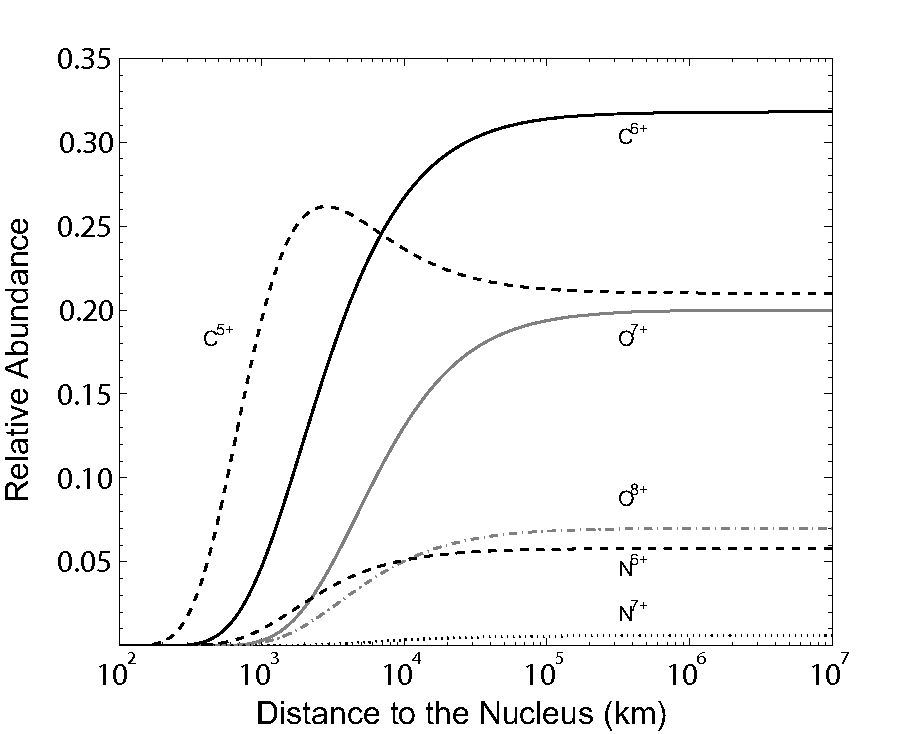}
   \caption{Modeled charge state distribution along the comet-Sun line, assuming an equatorial 300~km~s$^{-1}$ wind interacting with a comet with outgassing rate Q=$10^{29}$~\Q{} at 1~AU from the Sun. A composition typical for the slow, equatorial wind was assumed.}
              \label{fig:csd}%
    \end{figure}
%

Initially, the charge state distribution depends on the solar wind
state. For most simulation purposes, we will assume the `average'
ionic composition for the slow, equatorial solar wind as given by \cite{Sch00}. Using our compilation of charge changing cross
sections, we can solve the differential equations that describe
the charge state distribution in the coma in the 2D-geometry fixed
by the comet-Sun axis. Figure~\ref{fig:csd} shows the charge state
distribution for a 300~\kms{} equatorial wind interacting with a comet with an outgassing rate $Q$ of $=10^{29}$~\Q{} comet. From this charge state distribution,
it can be seen that along the comet-Sun axis, the comet becomes
collisionally thick between 3500~km (O$^{8+}$) to 2000~km
(C$^{6+}$), depending on the cross section of the ions. A maximum
in the C$^{5+}$ abundance can be seen around 2,000~km, which is
due to the relatively large initial C$^{6+}$ population and the
small cross section of C$^{5+}$ charge exchange.

A 3D integration assuming cylindrical symmetry around the
comet-Sun axis finally yields the absolute intensity of the
emission lines. Effects due to the observational geometry (i.e.
field of view and phase angle) are included at this step in the
model.


\section{Model Results}\label{sec:modelresults}

\subsection{Relative Contribution of Target Species}

   \begin{figure}
   \centering
    \includegraphics[width=7.5cm]{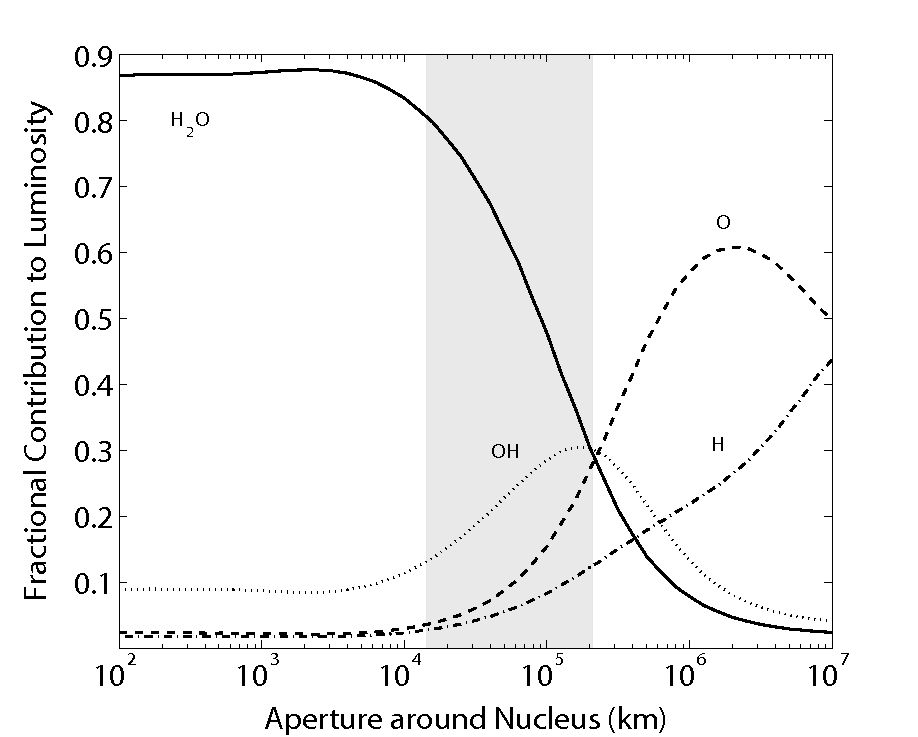}
   \caption{Relative contribution of target species to the total intensity of \ion{O}{vii} 570~eV emission
   complex with increasing field of view, for an active Q$=10^{29}$~\Q{} comet, interacting with a 300~\kms{}
   solar wind at 1~AU from the Sun. The shaded area indicates the range of apertures used to obtain spectra discussed within this survey.\label{fig:species}}
    \end{figure}

Figure~\ref{fig:species} shows the dominant collisions which
underly the X-ray emission of comets. Shown is the total
intensity projected on the sky, with increasing field of view.
Within $10^4$~km around the nucleus, water is the dominant collision partner. Farther outward ($\geq 2\times 10^5$~km), the
atomic dissociation products of water take over, and atomic oxygen
becomes the most important collision partner. When the field of view exceeds
10$^7$~km, atomic hydrogen becomes the sole collision partner.
Note that collisions with water never account for 100\% of the
emission, even with very small apertures, due to the contribution
of collisions with atomic hydrogen, OH and oxygen in the line of
sight towards the nucleus.

The comets observed with \emph{Chandra} are all observed with an
aperture of ca. 7.5\arcmin\ centered on the nucleus. This
corresponds to a range of $1.6 - 22 \times 10^4$~km (as indicated
in Figure~\ref{fig:species}). Our model predicts that the emission
from nearby comets will be dominated by \textsc{cxe} from water, but that
for comets observed with a larger field of view, up to 60\% of the
emission can come from \textsc{cxe} interactions with the water
dissociation products atomic oxygen and OH, and 10\% from
interactions with atomic hydrogen.

\subsection{Solar Wind Velocity}
\begin{figure}
\centering
\includegraphics[width=8cm]{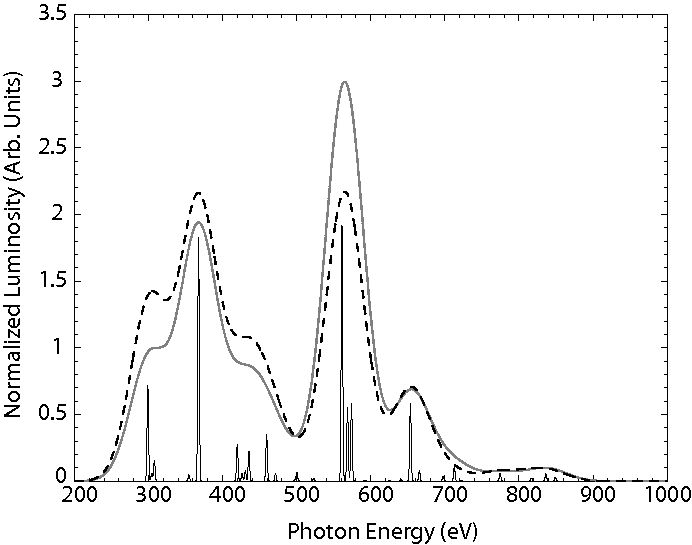}
\caption{Simulated X-ray spectra for a 10$^{29}$~\Q{} comet
interacting with an equatorial wind with velocities of 300~\kms{} (solid grey line) and 700~\kms{} (dashed black
line). The spectra are convolved with Gaussians with a width of
$\sigma=50$~eV to simulate the \emph{Chandra} spectral resolution.
To indicate the different lines, also the 700~\kms{} $\sigma=1$~eV
spectrum is indicated (not to scale). A field of view of 10$^5$~km
and `typical' slow wind composition were used.}
\label{fig:spectraV}
\end{figure}

To illustrate solar wind velocity induced variations in charge
exchange spectra, we simulated charge exchange spectra following
solar wind interactions between an equatorial wind and a $Q =
10^{29}$~\Q{} comet, and assumed the same solar wind composition
in all cases. In Fig.~\ref{fig:spectraV}, spectra resulting from
collisional velocities of 300~\kms{} and 700~\kms{} are
shown. In the spectrum from the faster wind, the \ion{C}{vi} 367~eV and \ion{O}{vii} 570~eV emission features are roughly equally
strong, whereas at 300~\kms, the oxygen feature is clearly
stronger. Assuming the wind's composition remains the same, within
the range of typical solar wind velocities (300--700~\kms), the
cross sectional dependence on solar wind velocity does not affect
cometary X-ray spectra by more than a factor 1.5. In practice,
the compositional differences between slow and fast wind will
induce much stronger spectral changes.

\subsection{Collisional Opacity}
   \begin{figure}
   \centering
    \includegraphics[width=7.5cm]{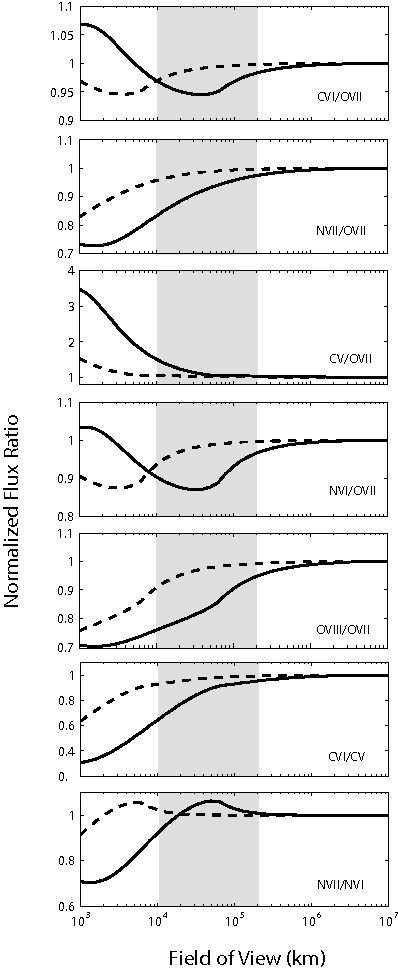}
   \caption{Collisional opacity effects on flux ratios within the field of view. The outer bounds of the fields of view within this survey were between $10^4-10^5$~km, as indicated by the shaded area. We considered a 500~\kms{} equatorial wind interacting with comets with different activities: $Q=10^{28}$~\Q\  (dashed lines) and $Q=10^{29}$~\Q{} (solid lines). All flux ratios are normalized to 1 at infinity.}
              \label{fig:ratios_model}%
    \end{figure}
%

Many of the 20$^+$ comets that have been observed in X-ray
display a typical crescent shape as the solar wind ion content is
depleted via charge exchange. Comets with low outgassing rates
around 10$^{28}$~\Q, such as 2P/2002~(Encke) and 9P/2005~(Tempel~1), did not display this emission morphology \cite{Lis05,
Lis07}. Whether or not the crescent shape can be resolved depends
mainly on properties of the comet (outgassing rate), but, to a
minor extent, also on the solar wind (velocity dependence of cross
sections). Other parameters (secondary, but important), are the
spatial resolution of the instrument and the distance of the comet
to the observer.

In a collisionally thin environment, the ratio between emission
features is the product of the ion abundance ratios and the ratio
between the relevant emission cross sections:
\begin{equation}\label{eq:thin}
r_{\mathrm{thin}} =
\frac{n(A^{q+})}{n(B^{q+})}\cdot\frac{\sigma_{\mathrm{em}}^{A^{q+}}(\mathrm{v})}{\sigma_{\mathrm{em}}^{B^{q+}}(\mathrm{v})}
\end{equation}

The flux ratio for a collisionally thick system depends on the
charge states considered. In case of a bare ion $A$ and a hydrogenic ion $B$, the
ratio between the photon fluxes from $A$ and $B$ is given by the
abundance ratio weighted by efficiency factors $\mu$ and $\eta$:

\begin{equation}\label{eq:thick}
r_{\mathrm{thick}} = \frac{n(A^{q+})}{n(B^{(r-1)+})+
\mu(B^{r+}) n(B^{r+})}\cdot \frac{\eta(A^{q+})}{\eta(B^{(r-1)+})}
\end{equation}

\noindent The efficiency factor $\mu$ is a measure of how much B$^{(r-1)+}$ is produced by charge exchange reactions by B$^{q+}$:
\begin{equation} \mu = \frac{\sigma_{r,r-1}(\mathrm{v})}{\sigma_r(\mathrm{v})} \end{equation} \noindent where $\sigma_r$ is the total charge exchange cross section and $\sigma_{r,r-1}$ the one electron charge changing cross section. The efficiency factor $\eta$ describes the emission yield per reaction and is given by the ratio between the
relevant emission cross section $\sigma_{\mathrm{em}}$ and the total charge
changing cross section $\sigma_r$:
\begin{equation}
\eta = \frac{\sigma_{\mathrm{em}}(\mathrm{v})}{\sigma_{r}(\mathrm{v})}
\end{equation}

To explore the effect of collisional opacity on spectra, we
simulated two comets at 1~AU from the Sun, with gas production rates
of $10^{28}$ and $10^{29}$~\Q, interacting with a solar wind with
a velocity of 500~\kms{} and an averaged slow wind composition
\cite{Sch00}. The results are summarized in
Figure~\ref{fig:ratios_model} where different flux ratios are
shown. The behavior of these ratios as a function of aperture is
important because they can be used to derive relative ionic
abundances. All ratios are normalized to 1 at infinite distance
from the comet's nucleus. For low activity comets with $Q \leq
10^{28}$~\Q, the collisional opacity does not affect the comet's
X-ray spectrum. Within typical field of views all line flux ratios
are close to the collisionally thin value. For more
active comets ($Q=10^{29}$~\Q), collisional opacity can become
important within the field of view. Observed flux ratios involving \ion{C}{v}
should be treated with care, see e.g. \ion{C}{v}/\ion{O}{vii} and
\ion{C}{vi}/\ion{C}{v}, because the flux ratios within the field of
view can be affected by almost 50\% and 35\%, respectively. The
effect is the strongest in these cases because of the large
relative abundance of C$^{6+}$, that contributes to the \ion{C}{v}
emission via sequential electron capture reactions in the
collisionally thick zones. For \ion{N}{vii} and \ion{O}{viii}, a
small field of view of $10^4$~km could affect the observed ionic ratios by
some 20\%.

\begin{figure}
\centering
\includegraphics[width=8cm]{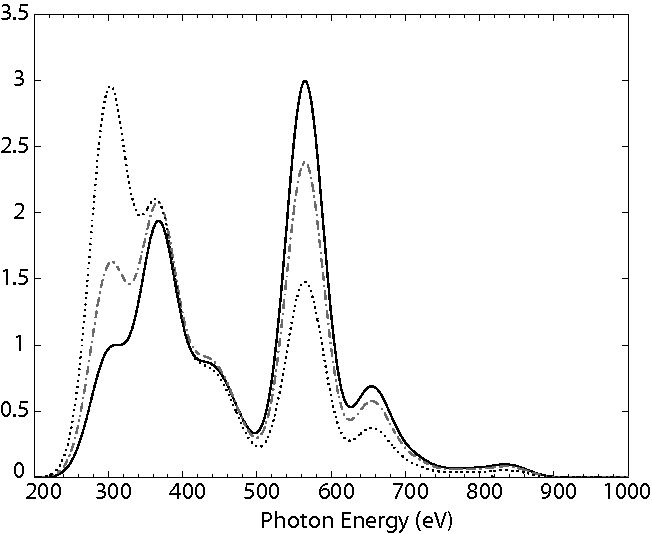}
\caption{Simulated X-ray spectra for a 10$^{29}$~\Q{} comet
interacting with an equatorial wind with a velocity of 300~\kms{} for fields of view decreasing
from 10$^5$~km (solid line), 10$^4$~km (dashed line) and 10$^3$~km (dotted line).}
\label{fig:spectraFOV}
\end{figure}

To further illustrate these results, we show the resulting X-ray
spectra in Fig.~\ref{fig:spectraFOV}. There, we consider a $Q =
10^{29}$~\Q{} comet interacting with a 300~\kms{} wind and show
the effect of slowly zooming from the collisionally thin to the
collisionally thick zone around the nucleus. The field of view
decreases from $10^5$ to $10^3$ km. At $10^5$~km, the spectrum is not affected by collisionally thick emission, whereas
the emission within an aperture of 1000~km is almost purely from
the interactions within the collisionally thick zones of the
comet, which can be most clearly seen by the strong enhancement of
the \ion{C}{v} emission around 300~eV.

The results of our model efforts demonstrate that cometary X-ray spectra reflect characteristics of the comet, the solar wind and the observational conditions.
Firstly, charge exchange cross sections depend on the velocity of the solar wind, but its effects are the strongest at velocities below regular solar wind velocities.
Secondly, collisional opacity can affect cometary X-ray spectra
but mainly when an active comet ($Q=10^{29}$~\Q)
is observed with a small field of view ($\leq 5\times 10^4$~km). The dominant factor however to explain differences in cometary \textsc{CXE} spectra is therefore the state and hence composition of the solar wind. This implies that the spectral analysis of cometary X-ray spectra can be used as a direct, remote quantitative and qualitative probe of the solar wind.


\section{Observations}\label{sec:obs}

\begin{sidewaystable*}
\begin{minipage}[t][180mm]{\textwidth}

\caption{Comet observation times and observing parameters. For
comets McNaught--Hartley and Ikeya--Zhang, the parameters were
averaged over the observation time span. Solar wind proton velocities and fluxes are measured in the ecliptic plane. Solar wind data for comets observed at large heliocentric latitudes are therefore highly uncertain and denoted within brackets. Comet labels refer to labels used in the figures in this paper.
  \label{tab:obs}} \centering
\begin{tabular}{l l c c c c c c c c} \hline\hline

Parameter   &       &   C/1999 S4   &   C/1999 T1   &   C/2000 WM1  &   153P/2002   &    2P/2003    &   C/2001 Q4   &   9P/2005 &   73P/2006    \\
    &       &   (\textsc{linear})    &   (McNaught--Hartley)  &   (\textsc{linear})    &   (Ikeya--Zhang)   &   (Encke) &   (\textsc{neat})  &   (Tempel~1)  &   (SW3-B) \\
\hline
\multicolumn{2}{l}{Comet Label} & a. & b. & c. &d. & e. & f. & g. & h.\\
\multicolumn{2}{l}{Observation Date}            &   7/14/2000    &   1/8-15/2001  &   12/31/2001    &   4/15-16/2002 &   12/24/2003    &   5/12/2004    &   6/30/2005   &   5/23/2006    \\
T$_{\textrm{exp}}$   &   (ksec)  &   9.4 &   16.9    &   35  &   24  &   54  &   10.51   &   52  &   20    \\
r$_h$\footnote{JPL-Horizons}    &   (AU)    &   0.80    &   1.26    &   0.75    &   0.81    &   0.89    &   0.96    &   1.51    &   0.97    \\
$\Delta^a$  &   (AU)    &   0.53    &   1.37    &   0.68    &   0.45    &   0.28    &   0.38    &   0.88    &   0.10    \\
Lon$^a$ &   (Degrees)   &   312 &   185 &   70  &   206 &   51  &   211 &   245 &   247 \\
$\Delta $Lon\footnote{difference between heliocentric longitude of comet and earth Lon$_c$ - Lon$_e$}   &   (Degrees)   &   20  &   73  &   -30 &   0.5 &   -11 &   -22 &   -34 &   5   \\
Lat$^a$ &   (Degrees)   &   24  &   15  &   -34 &   26  &   11  &   -3  &   0.8 &   0.5 \\
Phase$^a$   &   (Degrees)   &   98  &   62  &   87  &   102 &   103 &   86  &   41  &   114 \\
S-O-T   &   (Degrees)   &   51  &   44  &   49  &   52  &   60  &   72  &   105 &   61  \\
Q$_{\textrm{gas}}$   &   (10$^{28}$~mol. s$^{-1}$)  &   3\footnote{\citet{Boc01, Far01}}  &   6-20\footnote{\citet{Mum01, Sch01, Wea02, Biv06}}   &   3-9\footnote{ \citet{Sch02,Biv06}}   &   20\footnote{\citet{Del04,Biv06}}  &   0.7\footnote{\citet{Lis05}}   &   13\footnote{\citet{Fri05} and references therein}    &   0.9\footnote{\citet{Mum05,Sch06}} &   2\footnote{Schleicher, priv.~communication} \\
Wind    &       &   \textsc{icme}    &   \textsc{cir}/flare   &   \textsc{icme}    &   \textsc{icme}    &   Flare/PS\footnote{PS - post shock flow} &   Quiet   &   Quiet   &   \textsc{cir} \\
v$_p$\footnote{\textsc{ace-swepam} and \textsc{soho-celias} online data archives} &   (km s$^{-1})$   &   (592)   &   353 &   (324)   &   (372)   &   583   &   352 &   402 &   449 \\
$\Delta t$  &   (days)  &   1.09    &   6.63    &   -3.5    &   -0.73   &   -1.09   &   -1.80   &   -0.38   &   0.2 \\
F$_p^l$\footnote{The proton flux was scaled by 1/r$^2_h$ at the location of each comet to account for its density fall off due to radial expansion of the solar wind.}     &   (10$^8$~cm$^{-2}$ s$^{-1})$    &   (2.9) &   1.6    &   (1.4)    &   (3.8)    &   3.15   &   5.9    &   5.1    &   2.3  \\
Radius FoV  &   (10$^{5}$ km)   &   0.86    &   2.24    &   1.66    &   0.73    &   0.46    &   0.62    &   1.44    &   0.16    \\
\hline
\end{tabular}

\end{minipage}
\end{sidewaystable*}

In this section, we will briefly introduce the different comet
observations performed with \emph{Chandra}. A summary of comet and
solar wind parameters is given in Table~\ref{tab:obs}. More
observational details on the comet and a summary of the state of
the solar wind at the location of the comet during the X-ray
observations can be found in Appendix~\ref{sec:appendix}.

\subsection{Solar Wind Data}
   \begin{figure}
   \centering
    \includegraphics[width=7.5cm]{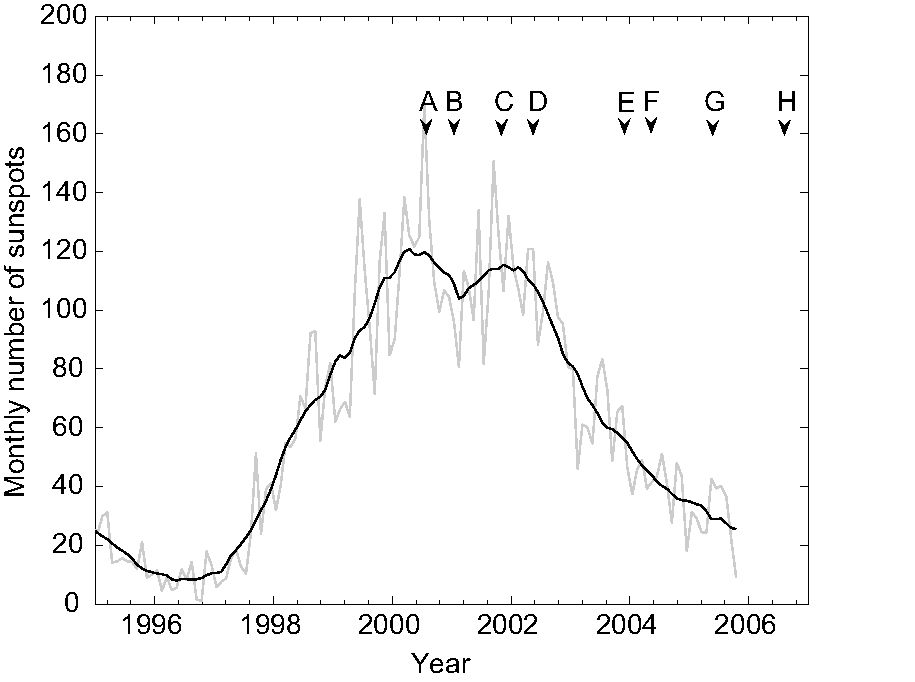}
   \caption{Chandra comet observations during the descending phase of solar cycle \# 23. Monthly sunspot numbers (grey line) and smoothed monthly sunspot number (black lines) from the Solar Influences Data Analysis Center of the Department of Solar Physics, Royal Observatory of
   Belgium (http://sidc.oma.be/). Letters refer to the chronological order of observation.\label{fig:cycle}}
    \end{figure}

\begin{figure*}
    \centering
        \includegraphics[width=0.9\textwidth]{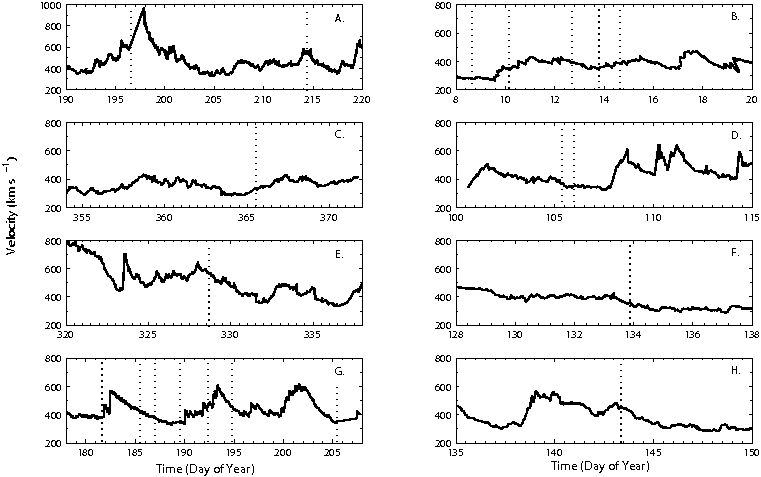}
    \caption{Solar wind proton velocities estimated from \textsc{ace} and \textsc{soho} data. For all
    comets, the time of the observations is indicated with a
    dotted line. Letters refer to the chronological order of observation.
\label{fig:sw_panel}}
\end{figure*}

Our survey spans the whole period between solar maximum (mid 2000)
and solar minimum (mid 2006), see Fig.~\ref{fig:cycle}. During
solar minimum, the solar wind can be classified in polar- and
equatorial streams, where the polar can be found at latitudes
larger than 30$^{\circ}$ and the equatorial wind within 15$^{\circ}$ of the
helioequator. Polar streams are fast (ca. 700~\kms) and
show only small variations in time, in contrast to the irregular
equatorial wind. Cold, fast wind is also ejected from coronal
holes around the equator, and when these streams interact with the
slower background wind corotating interaction regions (\textsc{cir}s) are
formed. As was illustrated by \citet{Sch00}, different wind types
vary greatly in their compositions, with the cooler, fast wind
consisting of on average lower charged ions than the hotter
equatorial wind. This clear distinction disappears during solar
maximum, when at all latitudes the equatorial type of wind dominates. In
addition, coronal mass ejections are far more common around solar
maximum.

There is a strong variability of heavy ion densities due to
variations in the solar source regions and dynamic changes in the
solar wind itself \cite{Zur06}. The variations mainly concern the
charge state of the wind as elemental variations are only on the
order of a factor of 2 (\citet{Von00}, and references therein).

We obtained solar wind data from the online data archives of \textsc{ace}
(proton velocities and densities from the \textsc{swepam}~instrument, heavy
ion fluxes from the \textsc{swics} and \textsc{swims}
instruments\footnote{http://www.srl.caltech.edu/ace/ASC/level2/index.html})
and \textsc{soho} (proton fluxes from the Proton Monitor
Instrument\footnote{http://umtof.umd.edu/pm/crn/}). Both \textsc{ace} and
\textsc{soho} are located near Earth, at its Lagrangian point L1. In order
to map the solar wind from L1 to the position of the comets, we
used the time shift procedure described by \citet{Neu00}. The
calculations are based on the comet ephemeris, the location of L1
and the measured wind speed. With this procedure, the time delay
between an element of the corotating solar wind arriving at L1 and
the comet can be predicted. A disadvantage of
this procedure is that it cannot account for latitudinal
structures in the wind or the magnetohydrodynamical behavior of
the wind (i.e., the propagation of shocks and \textsc{cme}s). These
shortcomings imply that especially for comets that have large
longitudinal, latitudinal and/or radial separations from Earth,
the solar wind data is at best an estimate of the local wind
conditions. The resulting proton velocities at the comets near the
time of the \emph{Chandra} observations are shown in
Fig.~\ref{fig:sw_panel}.

Parallel to this helioradial and heliolongitudinal mapping, we
compared our comet survey to a 3D \textsc{mhd} time--dependent solar wind
model that was employed during most of Solar Cycle 23 (1997 -
2006) on a continuous basis when significant solar flares were
observed. The model (reported by \cite{Fry03,McK06} and Z.K. Smith, private communication, for, respectively, the
ascending, maximum, and descending phases) treats solar flare
observations and maps the progress of interplanetary shocks and
\textsc{cme}s. The papers mentioned above provide an \textsc{rms} error for "hits"
of $\pm11$ hours \cite{Smi00, McK06}. \textsc{cir} fast forward shocks
were also taken into account in order to differentiate between the
co-rotating "quiet" and transient structures.  It was important,
in this differentiating analysis, to examine (as we have done
here) the ecliptic plane plots of both of these structures as
simulated by the deforming interplanetary magnetic field lines
(see, for example, \cite{Lis05,Lis07} for several of the comets
discussed here.) Therefore, the various comet locations
(Table~\ref{tab:obs}) were used to estimate the probability of
their X-ray emission during the observations being influenced by either of these heliospheric situations.

\subsection{X-ray Observations}
After its launch in 1999, 8 comets have been observed with the
\emph{Chandra} X-ray Observatory and Advanced \textsc{ccd} Imaging
Spectrometer (\textsc{acis}). Here, we have mainly considered observations
made with the \textsc{acis}-S3 chip, which has the most sensitive low
energy response and for which the majority of comets were
centered. The \emph{Chandra's} \textsc{acis}-S instrument provides moderate
energy resolution ($\sigma \approx$ 50~eV) in the 300 to 1500~eV
energy range, the primary range for the relatively soft cometary
emission. All comets in our sample were re-mapped into
comet-centered coordinates using the standard Chandra Interactive
Analysis of Observations (\textsc{ciao}~v3.4) software `sso\_freeze'
algorithm.

Comet source spectra were extracted from the S3 chip with a
circular aperture with a diameter of 7.5$\arcmin$, centered on the
cometary emission. The exception was comet C/2001~Q4, which filled the chip and a
50\% larger aperture was used. \textsc{acis}' response matrices were used to model the instrument's effective
area and energy dependent sensitivity matrices were created for
each comet separately using the standard \textsc{ciao} tools.

Due to the large extent of cometary X-ray emission, and
\emph{Chandra}'s relatively narrow field of view, it is not
trivial to obtain a background uncontaminated by the comet and
sufficiently close in time and viewing direction. We extracted
background spectra using several techniques: spectra from the S3
chip in an outer region generally $>8\arcmin$, an available \textsc{acis}
S3 blank sky observation, and backgrounds extracted from the \textsc{S1~ccd}. For several comets there are still a significant number of
cometary counts in the outer region of the \textsc{S3~ccd}. Background
spectra taken from the S1 chip have the advantage of having been taken
simultaneous with the S3 observation and thus having the same space
environment as the S3 observation. In general the background
spectra were extracted with the same 7.5$\arcmin$ aperture as the
source spectra but centered on the S1 chip. For comet Encke, where
the S1 chip was off during the observation the background from the
outer region of the S3 chip was used. Comet C/2000~WM1~(\textsc{linear})
was observed with the Low-Energy Transmission Grating (\textsc{letg}) and \textsc{acis-S} array. For the latter, we
analyzed the zero-th order spectrum, and used a background
extracted from the outer region of the S3 chip. It is possible
that the proportion of incident X-rays diffracted onto the S3 chip
will vary with photon energy. Background-subtracted spectra generally have a signal-to-noise at 561~eV of at least 10, and over 50 for 153P/2002 C1~(Ikeya--Zhang).

\section{Spectroscopy}\label{sec:fits}

\begin{figure}
    \centering
    \includegraphics[width = 8.0cm]{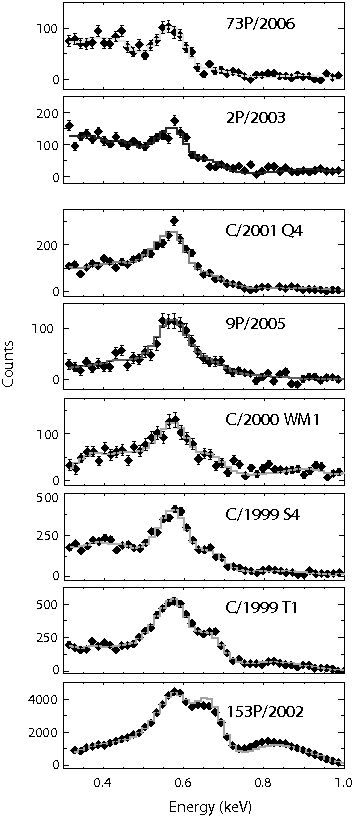}
    \caption{Observed spectrum and fit of all 8 comets observed
    with \emph{Chandra}, grouped by their spectral shape (see text). The histogram lines indicate the CXE model fit.
\label{fig:panel}}
\end{figure}

The observed spectra are shown in Figure~\ref{fig:panel}. The spectra suggest a classification based upon three competing emission features, i.e. the combined carbon and nitrogen emission (below~500~eV), \ion{O}{vii} emission around 565~eV and \ion{O}{viii} emission at 654~eV.
Firstly, the C+N emission ($<$500~eV) seems to be anti-correlated with the oxygen emission. This clearly sets the spectra of 73P/2006~S.--W.3B and 2P/2003~(Encke) apart, as for those two comets the C+N features are roughly as strong as the \ion{O}{vii} emission. In the spectra of
the remaining five comets, oxygen emission dominates over the carbon and nitrogen emission below 500~eV. The \ion{O}{viii}/\ion{O}{vii} ratio can be seen to increase continuously, culminating in the spectrum of 153P/2002~(Ikeya--Zhang) where the spectrum is completely dominated by oxygen emission with almost comparable \ion{O}{viii} and \ion{O}{vii} emission features.
From our modelling, we expect that the separate classes reflect different states of the solar wind, which imply different ionic abundances. To explore the obtained spectra more
quantitatively, we will use a spectral fitting technique based on
our \textsc{cxe} model to extract X-ray line fluxes.

\subsection{Spectral Fitting}
\begin{figure}
    \centering
        \includegraphics[height = 8cm]{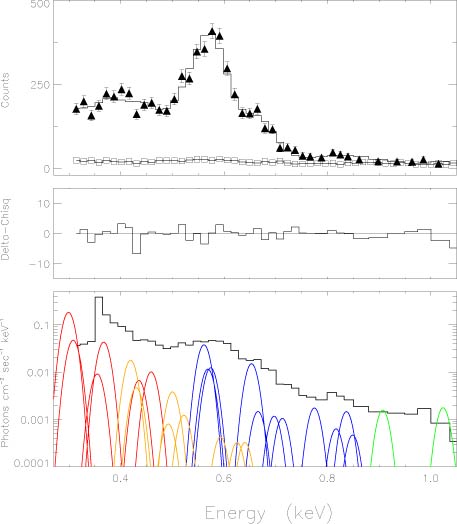}
    \caption{Details of the \textsc{cxe} fit for the spectrum of comet
    1999/S4 (\textsc{linear}). \textbf{Top panel:} Comet (filled triangles) and background (open squares)
    spectrum. \textbf{Middle panel:} Residuals of \textsc{cxe} fit \textbf{Bottom
    panel:} \textsc{cxe} model and observed spectrum
    indicating the different lines and
    their strengths. Carbon - red; nitrogen - orange; oxygen -
    blue; neon - green. The unfolded model is scaled above the
    emission lines for the ease of presentation.
\label{fig:LS4}}
\end{figure}


The charge exchange mechanism implies that cometary X-ray spectra result from a set of solar wind ions, which produce at least 35 emission lines in the regime visible with \emph{Chandra}. As comets are extended sources, these lines cannot all be resolved. All spectra were therefore fit using the 6 groups of fixed lines of our \textsc{cxe} model (see
Table~\ref{tab:em_cross}) and spectral parameters were derived
using the least squares fitting procedure with the \textsc{xspec} package.
The relative strengths from all lines were fixed per ionic species, according to their
velocity dependent emission cross sections. Thus,
the free parameters were the relative fluxes of the C, N and O ions contained in our model.

Two additional Ne lines at 907~eV (\ion{Ne}{ix}) and 1024~eV
(\ion{Ne}{x}) were also included, giving a total of 8 free
parameters. All line widths were fixed at the \textsc{acis-S3} instrument
resolution.

The spectra were fit in the 300 to 1000~eV range. This provided 49
spectral bins, and thus 41 degrees of freedom. \textsc{acis} spectra
below 300~eV are discarded because of the rising background
contributions, calibration problems and a decreased effective area
near the instrument's carbon edge.

As a more detailed example of the \textsc{cxe} model and comparison to the
data, we show in Fig~\ref{fig:LS4} the \textsc{acis-S3} data for C/1999 S4
(\textsc{linear}). The figure shows the background subtracted source
spectrum over-plotted with the background spectrum, the difference
between the model and data, and the model spectrum and data to indicate to contribution of the different ions.
Only the emission lines with $>$3\% strength of the strongest line
in their species are shown for ease of presentation.

The fluxes obtained by our fitting are converted into relative ionic abundances by
weighting them by their velocity dependent emission cross sections. For comets observed near the ecliptic plane ($<15^{\circ}$), solar wind conditions mapped to the comet were used (Section~4.1). For comets observed at higher latitudes, these data are most likely not applicable and a solar wind velocity of 500~\kms{} was assumed.

\begin{sidewaystable*}
\begin{minipage}[t][180mm]{\textwidth}

\caption[]{Results of the \textsc{cxe}-model fit for all cometary spectra.
Fluxes are given in dimensions of 10$^{-5}$~photons~cm$^{-2}$~\ps.
Errors are obtained by averaging over the calculated + and -
90$\%$ confidence contours and averaged those (corresponding to
$\chi^2=2.7$ or $1.6~\sigma$) \label{tab:fit}}\centering

\begin{tabular} {ll r@{\hspace{0.7mm}}c@{\hspace{0.7mm}}l r@{\hspace{0.7mm}}c@{\hspace{0.7mm}}l r@{\hspace{0.7mm}}c@{\hspace{0.7mm}}lr@{\hspace{0.7mm}}c@{\hspace{0.7mm}}lr@{\hspace{0.7mm}}c@{\hspace{0.7mm}}lr@{\hspace{0.7mm}}c@{\hspace{0.7mm}}lr@{\hspace{0.7mm}}c@{\hspace{0.7mm}}lr@{\hspace{0.7mm}}c@{\hspace{0.7mm}}lr@{\hspace{0.7mm}}c@{\hspace{0.7mm}}l}
\hline\hline
Ion &   Line    &   \multicolumn{3}{c}{C/1999 S4}           &   \multicolumn{3}{c}{C/1999 T1 }          &   \multicolumn{3}{c}{C/2000 WM1}          &   \multicolumn{3}{c}{153P /2002}          &   \multicolumn{3}{c}{2P/2003}         &   \multicolumn{3}{c}{C/2001 Q4}           &   \multicolumn{3}{c}{9P/2005}         &   \multicolumn{3}{c}{73P/2006}            \\  
    &   (eV)   &   \multicolumn{3}{c}{(\textsc{linear})}           &   \multicolumn{3}{c}{(McNaught--Hartley)}         &   \multicolumn{3}{c}{(\textsc{linear})}           &   \multicolumn{3}{c}{(Ikeya--Zhang)}          &   \multicolumn{3}{c}{(Encke)}         &   \multicolumn{3}{c}{(\textsc{neat})}         &   \multicolumn{3}{c}{(Tempel 1)}          &   \multicolumn{3}{c}{(SW3--B)}            \\  
\hline                                                                                                                                              
                                                                                                                                                
\ion{O}{viii}   &   653 &   23  &$\pm$& 2.0 &   25  &$\pm$& 1.6 &   20  &$\pm$& 3.9 &   357 &$\pm$& 14  &   1.95    &$\pm$& 0.36    &   9.7 &$\pm$& 1.7 &   3.5 &$\pm$& 0.98    &   1.5 &$\pm$& 0.57    \\  
\ion{O}{vii}    &   561 &   57  &$\pm$& 3.4 &   47  &$\pm$& 2.7 &   52  &$\pm$& 7.1 &   296 &$\pm$& 1.0 &   7.12    &$\pm$& 0.84    &   52  &$\pm$& 4.3 &   16  &$\pm$& 1.6 &   11  &$\pm$& 1.3 \\  
\ion{C}{vi} &   368 &   65  &$\pm$& 17  &   36  &$\pm$& 15  &   85  &$\pm$& 40  &   288 &$\pm$& 1.8 &   18  &$\pm$& 7.0 &   47  &$\pm$& 36  &   0.5 &$\pm$& 7.0 &   21  &$\pm$& 10  \\  
\ion{C}{v}  &   299 &   276 &$\pm$& 85  &   278 &$\pm$& 90  &   62  &$\pm$& 193 &   1052    &$\pm$& 18  &   192 &$\pm$& 62  &   809 &$\pm$& 280 &   121 &$\pm$& 89  &   326 &$\pm$& 109 \\  
\ion{N}{vii}    &   500 &   5.2 &$\pm$& 4.7 &   11  &$\pm$& 3.6 &   9.8 &$\pm$& 9.2 &   98  &$\pm$& 1.6 &   1.3 &$\pm$& 1.2 &   17  &$\pm$& 6.4 &   0   &$\pm$& 0.98    &   0.1 &$\pm$& 1.25    \\  
\ion{N}{vi} &   420 &   27  &$\pm$& 8.5 &   17  &$\pm$& 6.9 &   19  &$\pm$& 19  &   123 &$\pm$& 0.84    &   4.8 &$\pm$& 3.3 &   36  &$\pm$& 16  &   7.6 &$\pm$& 5.0 &   13.0    &$\pm$& 6.0 \\  
\ion{Ne}{ix}    &   907 &   2.4 &$\pm$& 1.1 &   1.97    &$\pm$& 0.67    &   6.7 &$\pm$& 1.7 &   38  &$\pm$& 0.11    &   0.68    &$\pm$& 0.2 &   1.4 &$\pm$& 0.54    &   0   &$\pm$& 0.17    &   0.48    &$\pm$& 0.24    \\  
\ion{Ne}{x} &   1024    &   2.7 &$\pm$& 1.2 &   0.41    &$\pm$& 0.47    &   4.6 &$\pm$& 2.1 &   8.2 &$\pm$& 0.02    &   0.72    &$\pm$& 0.24    &   0.8 &$\pm$& 0.46    &   0.07    &$\pm$& 0.22    &   0.02    &$\pm$& 0.25    \\  
 \hline                                                                                                                                             
$\chi^2_R$  &       &   \multicolumn{3}{c}{1.4}         &   \multicolumn{3}{c}{1.1}         &   \multicolumn{3}{c}{1.4}         &   \multicolumn{3}{c}{1.7}         &   \multicolumn{3}{c}{1.4}         &   \multicolumn{3}{c}{1.4}         &   \multicolumn{3}{c}{0.91}            &   \multicolumn{3}{c}{1}           \\  
\hline
\end{tabular}

\caption[]{Solar wind abundances relative to O$^{7+}$, obtained
from the \textsc{cxe}-model fit. \label{tab:ions}}\centering
\begin{tabular}{lr@{\hspace{0.7mm}}c@{\hspace{0.7mm}}lr@{\hspace{0.7mm}}c@{\hspace{0.7mm}}lr@{\hspace{0.7mm}}c@{\hspace{0.7mm}}lr@{\hspace{0.7mm}}c@{\hspace{0.7mm}}lr@{\hspace{0.7mm}}c@{\hspace{0.7mm}}lr@{\hspace{0.7mm}}c@{\hspace{0.7mm}}lr@{\hspace{0.7mm}}c@{\hspace{0.7mm}}lr@{\hspace{0.7mm}}c@{\hspace{0.7mm}}lr@{\hspace{0.7mm}}c@{\hspace{0.7mm}}l} \hline\hline
Ion &           \multicolumn{3}{c}{C/1999 S4}           &   \multicolumn{3}{c}{C/1999 T1 }          &   \multicolumn{3}{c}{C/2000 WM1}          &   \multicolumn{3}{c}{153P /2002}          &   \multicolumn{3}{c}{2P/2003}         &   \multicolumn{3}{c}{C/2001 Q4}           &   \multicolumn{3}{c}{9P/2005}         &   \multicolumn{3}{c}{73P/2006}            \\  
    &           \multicolumn{3}{c}{(\textsc{linear})}           &   \multicolumn{3}{c}{(McNaught--Hartley)}         &   \multicolumn{3}{c}{(\textsc{linear})}           &   \multicolumn{3}{c}{(Ikeya--Zhang)}          &   \multicolumn{3}{c}{(Encke)}         &   \multicolumn{3}{c}{(\textsc{neat})}         &   \multicolumn{3}{c}{(Tempel 1)}          &   \multicolumn{3}{c}{(SW3--B)}            \\  
\hline                                                                                                                                              
O$^{8+}$    &           0.32    &$\pm$& 0.03    &   0.42    &$\pm$& 0.04    &   0.31    &$\pm$& 0.07    &   0.96    &$\pm$& 0.04    &   0.19    &$\pm$& 0.04    &   0.17    &$\pm$& 0.03    &   0.19    &$\pm$& 0.06    &   0.10    &$\pm$& 0.04    \\  
C$^{6+}$    &           1.4 &$\pm$& 0.4 &   0.95    &$\pm$& 0.4 &   2.0 &$\pm$& 1.0 &   1.21    &$\pm$& 0.0 &   2.9 &$\pm$& 1.1 &   1.2 &$\pm$& 0.9 &   0.0 &$\pm$& 0.6 &   2.3 &$\pm$& 1.1 \\  
C$^{5+}$    &           12  &$\pm$& 3.8 &   15  &$\pm$& 4.8 &   3.0 &$\pm$& 9.3 &   8.9 &$\pm$& 0.2 &   56  &$\pm$& 19.3    &   49  &$\pm$& 17.4    &   22  &$\pm$& 15.9    &   71  &$\pm$& 24.9    \\  
N$^{7+}$    &           0.07    &$\pm$& 0.06    &   0.19    &$\pm$& 0.06    &   0.14    &$\pm$& 0.14    &   0.25    &$\pm$& 0.00    &   0.14    &$\pm$& 0.13    &   0.25    &$\pm$& 0.10    &   0.00    &$\pm$& 0.05    &   0.008   &$\pm$& 0.08    \\  
N$^{6+}$    &           0.63    &$\pm$& 0.21    &   0.47    &$\pm$& 0.20    &   0.49    &$\pm$& 0.49    &   0.56    &$\pm$& 0.00    &   0.79    &$\pm$& 0.55    &   1.1 &$\pm$& 0.52    &   0.72    &$\pm$& 0.48    &   1.5 &$\pm$& 0.72    \\  
Ne$^{10+}$  &           0.02   &$\pm$& 0.01   &   0.004   &$\pm$& 0.005   &   0.044   &$\pm$& 0.02   &   0.01    &$\pm$& 0.0001   &   0.05   &$\pm$& 0.02   &   0.008   &$\pm$& 0.004   &   0.002   &$\pm$& 0.007   &   0.001   &$\pm$& 0.01   \\  
 \hline                                                                                                                                             

\end{tabular}
\end{minipage}
\end{sidewaystable*}

\begin{figure}
  \begin{center}
 \includegraphics[width = 8.0cm]{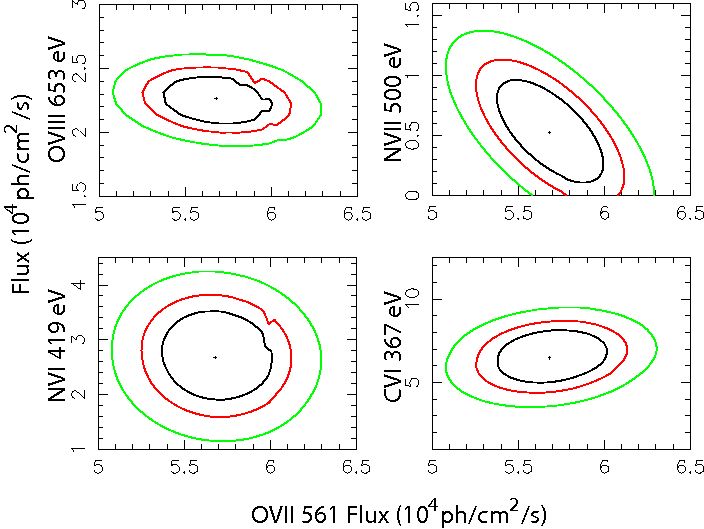}
     \caption{Parameter sensitivity for the major emission features in the fit of C/1999 S4 (\textsc{linear}), with respect to the \ion{O}{vii}~561~eV feature. All units are 10$^{-4}$~photons~cm$^{-2}$~\ps. The contours indicate a $\chi_R^2$ of 9.2 (or 99$\%$ confidence, largest, green contour), a $\chi_R^2$ of 4.6 (90$\%$, red contour) and a $\chi_R^2$ of 2.3 (68$\%$, smallest, blue contour).}
    \label{fig:contours}
  \end{center}
\end{figure}

\subsection{Spectroscopic Results}\label{sec:xrayspecresults}
\begin{table*}
\caption{Solar wind abundance relative to O$^{7+}$, obtained for
comet \textsc{linear}~S4. References: Bei '03 -- \citet{Bei03}, Kra
'04 -- \citet{Kra04}, Kra '06 -- \citet{Kra06}, Otr '06 --
\citet{Otr06} and S\&C '00 -- \citet{Sch00}. Dots indicate that an ion was included in the fitting, but no abundances were derived; dash means that an ion was not included in the fitting. \citet{Otr06} did not fit the observed spectrum, but used a combination of \textsc{ace}-data and solar wind averages from \protect \cite{Sch00} to compute a syntectic spectrum of the comet. Solar wind averages are given for comparison \citet{Sch00}\label{tab:ls4comp}} \centering
\renewcommand{\footnoterule}{}
\begin{tabular}{l r@{\hspace{0.7mm}}c@{\hspace{0.7mm}}l r@{\hspace{0.7mm}}c@{\hspace{0.7mm}}l r@{\hspace{0.7mm}}c@{\hspace{0.7mm}}l r@{\hspace{0.7mm}}c@{\hspace{0.7mm}}l c c} \hline\hline
Ion &   \multicolumn{3}{c}{this work}           &   \multicolumn{3}{c}{Bei 03}  &   \multicolumn{3}{c}{Kra 04}  &   \multicolumn{3}{c}{Kra 06}  &   Otr 06  \\

\hline                                                          \\
O$^{8+}$    &   0.32    &$\pm$& 0.03    &   0.13 &$\pm$& 0.03   &   0.13 &$\pm$& 0.05   &   0.15 &$\pm$& 0.03   &   0.35    &   0.35    \\
C$^{6+}$    &   1.4 &$\pm$& 0.4 &   0.9 &$\pm$& 0.3 &   0.7 &$\pm$& 0.2 &   0.7 &$\pm$& 0.2 &   1.02    &   1.59    \\
C$^{5+}$    &   12  &$\pm$& 4.0 &   11 &$\pm$& 9    &   \multicolumn{3}{c}{\ldots}  &   1.7 &$\pm$& 0.7 &   1.05    &   1.05    \\
N$^{7+}$    &   0.07    &$\pm$& 0.06    &   0.06 &$\pm$& 0.02   &   \multicolumn{3}{c}{--}  &   \multicolumn{3}{c}{--}  &   0.03    &   0.03    \\
N$^{6+}$    &   0.63    &$\pm$& 0.21    &   0.5 &$\pm$& 0.3 &   \multicolumn{3}{c}{--}  &   \multicolumn{3}{c}{--}  &   0.29    &   0.29    \\
Ne$^{10+}$  &   0.02    &$\pm$& 0.01    &   \multicolumn{3}{c}{--}  &   \multicolumn{3}{c}{--}  &   \multicolumn{3}{c}{--}  &   --  &   --  \\
Ne$^{9+}$   &   \multicolumn{3}{c}{\ldots}          &   \multicolumn{3}{c}{--}  &   \multicolumn{3}{c}{$(15 \pm 6) \times 10^{-3}$} &   \multicolumn{3}{c}{$(20 \pm 7) \times 10^{-3}$} &   --  &   --  \\
 \hline

\end{tabular}
\end{table*}

The fits to all cometary spectra are shown in
Fig.~\ref{fig:panel} and the results of the fits are given in
Table~\ref{tab:fit}. For the majority of the comets, the model is a good fit to the data within a 95\% confidence limit
($\chi_R^2 \approx 1.4$).
Results for comet 153P/2002~(Ikeya--Zhang) are presented in Table
5 with an additional systematic error to account for its brightness and any
uncertainties in the response.

The spectra for all comets are well reproduced in the 300 to
1000~eV range. The nitrogen contribution is
statistically significant for all comets except the fainter ones,
2P/2003~(Encke) and 73P/2006~(S.-W.3B). For example, removing the
nitrogen components from \textsc{linear}~S4's \textsc{cxe} model
and re-fitting, increases $\chi^2_R$ to over 7.

$\chi^2$ contours
for C/1999~S4~(\textsc{linear}) are presented in
Fig~\ref{fig:contours}. The line strengths for each ionic species are generally well
constrained, except where spectral features overlap. This can be readily seen when comparing the contours for the
\ion{N}{vii} 500~eV and \ion{O}{vii} 561~eV features where a
strong anti-correlation exists (Figure 12). Due to the limited
resolution of \textsc{acis} an increase in the \ion{N}{vii} feature will
decrease the \ion{O}{vii} strength. Similar anti-correlations
exist between the nitrogen \ion{N}{vi} or \ion{N}{vii} and
\ion{C}{v} 299~eV lines. Since the line strength for the main line
in each ionic species is linked to weaker lines, a range of
energies can contribute and better constrain its strength. However
with \ion{O}{vii} as the strongest spectral feature the nitrogen and carbon
components may be artificially lower as a result of the
aforementioned anti-correlations. The lack of effective area due
to the carbon edge in the \textsc{acis} response also may over-estimate the
\ion{C}{v} line flux. The neon features were well constrained for
the brighter comets, but this is a region of lower signal and some
caution must be taken when treating the neon line strengths and
they are included here largely for completeness.

In the case of 153P/Ikeya--Zhang, the $\chi^2_R > 1.4$. The main discrepancy is that the model produces
not enough flux in the 700 to 850~eV range compared to the
observed spectrum. This may reflect an underestimation of higher
 \ion{O}{viii} transitions or the presence of species not (yet) included in the model, such as Fe. This will be
discussed further in the last section of this paper and in a
separate paper dedicated to the observations of this comet
(K.~Dennerl, private communication).

One of the best studied comets is C/1999~S4~(\textsc{linear}), because of
its good signal-to-noise ratio. To discuss our results, we will
compare our findings with earlier studies of this comet. In general, the spectra analyzed here have more counts than
earlier analyzes, because of improvements in the \emph{Chandra}
processing software and because we took special care to use a
background that is as comet-free as possible. Previous studies
appear to have removed true comet signal when the background
subtraction was performed. In particular, both the \cite{Kra04}
and \cite{Lis01} studies used background regions from the outer
part of the S3 chip and this may have still had true cometary
emission. \cite{Kra04} subtracted over 70\% of the total signal as
background. We find that using the S1-chip, the background contributes only 20\% of
the total counts.

Different attempts to derive relative ionic abundances from C1999/S4's X-ray spectrum are compared in Table~\ref{tab:ls4comp}. Our atomic physics based spectral analysis combines the benefits of earlier analytical approaches by \citet{Kha00,Kha01,Bei03}. These methods were all applied to just one or two comets.
\citet{Bei03} interpret C1999/S4's X-ray spectrum by fitting it
with 6 experimental spectra obtained with their \textsc{ebit} setup. The resulting abundances are very similar to ours. The advantage of their method is that it
includes multiple electron capture, but in order to observe the
forbidden line emission, the spectra were obtained with trapped
ions colliding at CO$_2$, at collision energies of 200 to 300~eV
or ca. 30~\kms. As was shown in Fig.~\ref{fig:hardness}, the \textsc{cxe}
hardness ratio may change rapidly below 300~\kms, implying
an overestimation of the higher order lines compared to the $n =
2 \to 1 $ transition, which for \ion{O}{vii} overlap with the \ion{O}{viii} emission. We therefore find higher abundances of
O$^{8+}$.

\cite{Kra04,Kra06} obtained fluxes and ionic abundances by fitting the
spectrum with 10 lines of which the energies were semi-free. Their analysis thus does not
take the contamination of unresolved emission into account, and
\ion{N}{vi} and \ion{N}{vii} are not included in the fit. The
line energies were attributed to \textsc{cxe} lines of mainly solar wind C
and O but also to ions of Mg and Ne. The inclusion of the resulting low energy emission (near 300~eV) results in lower C$^{5+}$ fluxes (see also \citet{Otr06}).

There are several factors that may contribute to the unexpectedly low \ion{C}{vi}/\ion{C}{v} ratios: 1) There may be a small contribution to the \ion{C}{v} line from other ions in the 250-300~eV range (e.g. Si, Mg, Ne) that are currently not included in the model. Including these species in the model would lower the \ion{C}{v} flux, but probably only with a small amount. 2) The low \textsc{acis} effective area in the 250-300~eV region allows the \ion{C}{v} flux to be unconstrained, and this increases the uncertainty in the \ion{C}{v} flux. We estimate that the uncertainty in the effective area, introduced by the carbon edge, can account for an uncertainty as large as a factor of 10 in the observed \ion{C}{v}/\ion{C}{vi} ratios.

We will not compare our results with measured \textsc{ace/swics} ionic
data. As discussed in section~\ref{sec:obs}, the solar wind is
highly variable in time and its composition can change
dramatically over the course of less than a day. Variations in the
solar wind's ionic composition are often more than 50\% during the
course of an observation. Data on N, Ne, and O$^{8+}$ ions have
not been well documented as the errors of these abundances are
dominated by counting statistics. As discussed above, latitudinal
and corotational separations imply large inaccuracies in any solar
wind mapping procedure. These conditions clearly disfavor
modelling based on either average solar wind data or \textsc{ace/swics}
data.

\section{Comparative Results}\label{sec:discussion}
\begin{figure}
    \centering
    \includegraphics[width = 8.0cm]{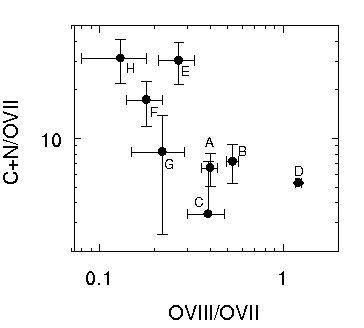}
    \caption{Flux ratios of all observed comets. The low energy C+N feature is anti-correlated to the oxygen ionic ratio. Letters refer to the chronological order of observation.\label{fig:flux_ratios}}
    \includegraphics[width = 8.0cm]{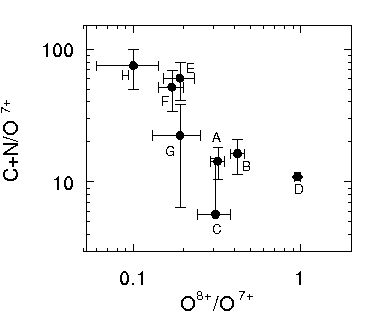}
    \caption{Ion ratios of all observed comets. The C+N ionic abundantie is anti-correlated to the oxygen ionic ratio. Letters refer to the chronological order of observation.
\label{fig:ion_ratios}}
\end{figure}
\begin{table*}
\caption{Correlation between classification according to spectral
shape and comet/solar wind characteristics during the
observations. Comet families from \cite{Mar05}. Phase refers to where in the solar cycle the comet was observed, where 1 is the solar maximum and 0 the solar minimum of cycle \#23's descending phase. For other
references, see Table~\ref{tab:obs}.\label{tab:parametertest}}
\centering
\renewcommand{\footnoterule}{}
\begin{tabular}{lc l r@{\hspace{0.7mm}}c@{\hspace{0.7mm}}l c c c c} \hline\hline
Class   &   \#  &   Comet       &   Comet   &   Q   &   Latitude    &      Wind Type   \\
    &       &           &   Family  &   (10$^{28}$ mol. s$^{-1}$)   &          &       \\
\hline
cold    &   H   &   73P/2006    (S.-W.3B)   &   Jupiter &   2   &   0.5        &   \textsc{cir}    \\
    &   E   &    2P/2003    (Encke) &   Jupiter &   0.7 &   11.4           &   Flare/PS    \\
\hline
warm    &   F   &   C/2001 Q4   (\textsc{neat}) &   \emph{unknown}  &   10  &   -3         &   Quiet   \\
    &   G   &   9P/2005 (Tempel 1)  &   Jupiter &   0.9 &   0.8 &  Quiet   \\
\hline
hot &   C   &   C/2000 WM1  (\textsc{linear})   &   \emph{unknown}  &   3-9 &   -34      &   PS  \\
    &   A   &   C/1999 S4   (\textsc{linear})   &   \emph{unknown}  &   3   &   24        &   \textsc{icme}   \\
    &   B   &   C/1999 T1   (McNaught--Hartley) &   \emph{unknown}  &   6-20    &   15  &   Flare/\textsc{cir}  \\
    &   D   &   C/2002 C1   (Ikeya--Zhang)  &   Oort    &   20  &   26  & \textsc{icme}   \\
\hline

\end{tabular}
\end{table*}

As noted in Section~\ref{sec:fits}, spectral differences show up in the behavior of the low energy C+N emission ($<$~500~eV), the \ion{O}{vii} emission at 561~eV and the \ion{O}{viii} emission at 653~eV. Figure~\ref{fig:flux_ratios} shows a color plot of the fluxes of these three emission features, and Figure~\ref{fig:ion_ratios} the corresponding ionic abundances. There is a clear separation between the two comets with a large C+N contribution and the other `oxygen-dominated' comets, which on their turn show a gradual increase in the oxygen ionic ratio. This sample of comet observations suggest that we
can distinguish two or three spectral classes.

Table~\ref{tab:parametertest} surveys the comet parameters for the
different spectral classes. The outgassing rate, heliocentric- or geocentric distance and comet
family do not correlate to the different classes, in accordance with our model findings. The data does suggest a correlation between latitude and wind conditions during the observations.
At first sight, the apparent correlation between latitude and oxygen ratio seems paradoxical. According to the bimodal structure of the solar wind the fast, cold wind dominates at latitudes $>15^{\circ}$, implying less \ion{O}{viii} emission. In Figure~\ref{fig:cycle}, the comet observations are shown with respect to the phase of the last solar cycle.
Interestingly, we note that all comets that were observed at
higher latitudes were observed around solar maximum. The solar wind is highly chaotic during solar maximum and the frequency of impulsive events like \textsc{CME}s is much higher than during the descending and minimum phase of the cycle. This explains both why the comets observed in the period 2000--2002 encountered a disturbed solar wind and
why our survey does not contain a sample of the cool fast wind from polar coronal holes.

The observed classification can therefore be fully ascribed to solar wind states. The first class is associated with cold, fast winds with
lower average ionization. These winds are found in \textsc{cir}s and behind
flare related shocks. The spectra due to these winds are dominated by the low energy x-rays, because of the low abundances of highly charged oxygen. At the relevant temperatures, most of the solar wind oxygen is He-like O$^{6+}$, which does not produce any emission visible in the 300--1000~eV regime accessible with \emph{Chandra}. Secondly, there is an intermediate class
with two comets that were all observed during periods of quiet
solar wind. These comets interacted with the equatorial, warm slow
wind. The third class then comprises comets that interacted with a
fast, hot, disturbed wind associated with \textsc{icme}s or flares. From the solar wind
data, Ikeya--Zhang was probably the most extreme example of this
case. This comet had 10 times more signal than any other comet in
our sample and small discrepancies in the response may be
important at this level. Extending into the 1-2~keV regime, a
preliminary analysis indicates the presence of bare and H-like Si,
Mg and \ion{Fe}{xv-xx} ions, in accordance with \textsc{ace} measurements
of \textsc{icme} compositions \cite{Lep04}.

   \begin{figure}
   \centering
    \includegraphics[width=7.5cm]{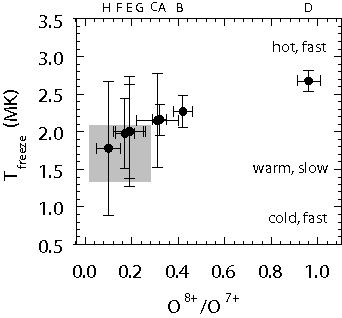}
   \caption{Spectrum derived ionic oxygen ratios and corresponding freezing-in temperatures from \cite{Maz98}. The shaded area indicates the typical range of slow wind associated with streamers. Letters refer to the chronological order of observation.\label{fig:OT}}
    \end{figure}

The variability and complex nature of the solar wind allows for
many intermediate states in between these three categories
\cite{Zur02}, which explain the gradual increase of the \ion{O}{viii}/\ion{O}{vii} ratio that we observed in the cometary spectra. As the solar wind is a collisionless plasma, the
charge state distribution in the solar wind is linked to the
temperature in its source region. Ionic temperatures are
therefore a good indicator of the state of the wind
encountered by a comet. The ratio between O$^{7+}$ and O$^{6+}$ ionic abundances
has been demonstrated to be a good probe of solar wind states.
\citet{Zur02} observed that slow, warm wind associated with streamers
typically lies within $0.1<$ O$^{7+}$/O$^{6+}$ $<1.0$, corresponding to freezing in temperatures of 1.3--2.1~MK. The
corresponding temperature range is indicated in the Figure~\ref{fig:OT}.
In the figure, we show the observed O$^{8+}$ to
O$^{7+}$ ratios and the corresponding freezing-in temperatures
from the ionizational/recombination equilibrium model by
\citet{Maz98}. Most observations are within or near to the
streamer-associated range of oxygen freezing in temperatures. Four comets interacted
with a wind significantly hotter than typical streamer winds, and in all four cases we found evidence in solar wind archives that the comets most likely encountered a disturbed wind.

\section{Conclusions}\label{sec:conclusions}

Cometary X-ray emission arises from collisions between bare- and
H-like ions (such as C, N, O, Ne, \ldots) with mainly water and
its dissociation products OH, O and H. The manifold of
dependencies of the \textsc{cxe} mechanism on characteristics of both comet
and wind offers many diagnostic opportunities, which are explored
in the first part of this paper. Charge exchange cross sections are strongly dependent on the velocity of the solar wind, and these effects are strongest at velocities below the regular wind conditions. This dependency might be used as a remote plasma diagnostics in future observations.
Ruling out collisional opacity effects, we used our model to demonstrate that the spectral shape of cometary
\textsc{cxe} emission is in the first place determined by local solar wind conditions. Cometary X-ray spectra hence reflect the state of the solar wind.

Based on atomic physic modelling of cometary charge exchange
emission, we developed an analytical method to study cometary
X-ray spectra. First, the data of 8 comets observed with \emph{Chandra} were carefully reprocessed to avoid the subtraction of cometary signal as background. The spectra were then fit using an extensive data set of velocity dependent emission cross sections for eight different solar wind species. Although the limited observational resolution currently available hampers the interpretation of
cometary X-ray spectra to some degree, our spectral analysis allows
for the unravelling of cometary X-ray spectra and allowed us to derive relative solar wind abundances from the spectra.

Because the solar wind is a collisionless plasma, local ionic charge states reflect conditions of its source regions. Comparing the fluxes of the C+N emission below 500~eV, the \ion{O}{vii} emission and the \ion{O}{viii} emission yields a quantitative probe of the
state of the wind. In accordance with our modelling, we found that spectral differences amongst the comets in our survey could be very well understood in terms of solar wind conditions. We are able
to distinguish interactions with three different wind types, being the cold, fast wind (I), the
warm, slow wind (II); and the hot, fast, disturbed wind (III). Based on our findings,
we predict the existence of even cooler cometary X-ray spectra
when a comet interacts with the fast, cool high latitude wind from
polar coronal holes. The upcoming solar minimum offers the perfect
opportunity for such an observation.


\begin{acknowledgements}
DB and RH acknowledge support within the framework of the
\textsc{fom--euratom} association agreement and by the Netherlands
Organization for Scientific Research (\textsc{nwo}). MD thanks the
\textsc{noaa} Space Environment Center for its post-retirement
hospitality. We are grateful for the cometary ephemerides of
D.~K.~Yeomans published at the \textsc{jpl/horizons} website.
Proton velocities used here are courtesy of the
\textsc{soho/celias/pm} team. \textsc{soho} is a mission of
international cooperation between \textsc{esa} and \textsc{nasa}. \textsc{Chianti} is a collaborative project involving the \textsc{nrl} (USA), \textsc{ral} (UK), \textsc{mssl} (UK), the Universities of Florence (Italy) and Cambridge (UK), and George Mason University (USA).

\end{acknowledgements}

\bibliographystyle{aa} 

\appendix
\section{Observations within this Survey} \label{sec:appendix}
This Appendix presents the observational details of the \emph{Chandra} data and the corresponding solar wind state. The prefix 'FF' (fearless forecast) used in this appendix refers
to the real time forecasting of coronal mass ejection shocks
arrivals at Earth. The numbers were so-named for flare/coronal
shock events during solar cycle \#23.

\subsection{C/1999~S4~(\textsc{linear})}
\noindent\textbf{X-rays.} The first {\em Chandra} cometary
observation was of comet C/1999~S4~(\textsc{linear}) \cite{Lis01}, with
observations being made both before and after the breakup of the
nucleus.  Due to the low signal-to-noise ratio of the second
detection, only the July 14th 2000 pre-breakup observation is
discussed here.  Summing the 8 pointings of the satellite gave a
total time interval of 9390~s. In this period, the \textsc{acis-S3 ccd}
collected a total of 11\,710~photons were detected in the range
300--1000~eV. Detections out side this range or on other \textsc{acis-ccd}s were not attributed to the comet.
As a result, data from the S1-\textsc{ccd} (which is configured identically
to S3) may be used as an indicator of the local X-ray background.

The morphology can be described by a crescent shape, with the maximum brightness point 24\,000~km from the nucleus on the Sun-facing side. The brightness dims to 10\% of the maximum level at 110\,000~km from the nucleus.\\

\noindent\textbf{Solar wind.} A large velocity jump can be seen
around DoY~199, which was due to the famous "Bastille Day" flare
on 14 July (FF\#153, \cite{Dry01,Fry03}). This flare reached the
comet only after the first observation. At July~12, 2017UT a solar
flare started at N17W65~(FF\#152), which was nicely placed to hit
this comet with a very high probability during the first
observations \cite{Fry03}. As for the second observation, there
was another flare on July 28, S17E24, at 1713 UT (FF\#164) and
there was a high probability that its shock's weaker flank hit the
comet.

\subsection{C/1999 T1 (McNaught--Hartley)}
\noindent\textbf{X-rays.} The allocated observing time of comet
McNaught--Hartley was partitioned into 5~one-hour-slots between
January 8th and January 15th, 2001 \cite{Kra02}. The strongest
observing period was on January 8th, when $\Delta = 1.37$ AU and
$r_h = 1.26$~AU.

There were 15\,000 total counts observed by the \textsc{acis-S3~ccd} between
300 and 1000~eV. The emission region can be described by a
crescent, with the peak brightness is at 29\,000~km from the
nucleus.  The brightness dims to 10\% of the maximum at a
cometocentric distance of 260\,000~km.
Again, the \textsc{acis-S1 ccd} may be used to indicate the local background signal.\\

\noindent\textbf{Solar wind.} The comet was not within the
heliospheric current/plasma sheet (\textsc{HCS/HPS}). Two corotating \textsc{cir}s
are probably associated with the first two observations. Two
flares (FF\#233 and \#234) took place; however, another corotating
\textsc{cir} more likely arrived before the flare's transient shock's
effects did \cite{McK06}.

\subsection{C/2000 WM1 (\textsc{linear})}
\noindent\textbf{X-rays.} The only attempt to use the
high-resolution grating capability of the \textsc{acis-S} array was made
with comet C/2000~WM1~(\textsc{linear}). Here, the Low-Energy Transmission
Grating (\textsc{letg}) was used. The dimness of the observed X-rays, and
the extended nature of the emitting atmosphere meant that the
grated spectra did not yield significant results. It is still
possible to extract a spectrum based on the pulse-heights
generated by each X-ray detection on the \textsc{acis-S3}~chip, although
the morphology is not recorded. 6300 total counts were recorded
for the pulse-height spectrum of the S3 chip in the 300 to 1000~eV
range.\\

\noindent\textbf{Solar wind.} Comet WM1 was observed at the
highest latitude available within this survey, and at a latitude
of 34 degrees, it was far outside the \textsc{hcs}. During the
observations, this comet might have experienced the southerly
flank of the shock of a strong X3.4 flare at S20E97 and its \textsc{icme}
and shock on December 28, 2001 (FF\#359) \cite{McK06}.

\subsection{153P/2002 (Ikeya--Zhang)}
\noindent\textbf{X-rays.} The brightest X-ray comet in the Chandra
archive is 153P/2002~(Ikeya--Zhang).  The heliographic latitude,
geocentric distance and heliocentric distance were comparable to
those for comet C/1999~S4~(\textsc{linear}), with a latitude of
$26^{\circ}$, $\Delta = 0.457$~AU and $r_h = 0.8$~AU. Rather than
periodically re-point the detector to track the comet, the
pointing direction was fixed and the comet was monitored as it
passed through the field of view, thus increasing the effective
FoV. There were two observing periods on April 15th 2002, each
lasting for approximately 3~hours and 15~minutes.  In both
periods, a strong cometary signal is detected on all of the
activated \textsc{acis-ccd}s. Consequently, a background signal cannot be
taken from the observation.  A crescent shape on the Sun side of
the comet is observed over all of the \textsc{ccd} array. Over 200\,000
total counts were observed from the S3 chip in the 300 to 1000~eV
range. The time intervals for each observing period are 11\,570 and
11\,813~seconds.\\

\noindent\textbf{Solar wind.} Like C/2000~WM1, this comet was
observed at a relatively high heliographic latitude. Solar wind
data obtained in the ecliptic plane can therefore not be used to
determine the wind state at the comet. 153P/2002~(Ikeya--Zhang) was
well-positioned during the first observation on 15 April 2002 for
a flare at N16E05 (FF\#388) on 12 April 2002. During the second
observation on 16 April, there was an earlier flare on 14 April at
N14W57, but this flare was probably too far to the west to be
effective \cite{McK06}. The comet was observed at a high latitude, and hence \textsc{ace} solar wind
data is most likely not applicable.

\subsection{2P/2003 (Encke)}
\noindent\textbf{X-rays.} The Chandra observation of Encke took
place on the 24th of November 2003 \cite{Lis05}, when the comet
had a heliocentric distance of $r_h = 0.891$~AU and a geocentric
distance of $\Delta = 0.275$~AU and a heliographic latitude of
11.4 degrees. The comet was continuously tracked for over 15
hours, resulting in a useful exposure of 44\,000~seconds. The
\textsc{acis-S3~ccd} counted 6140~X-rays in the range 300--1000~eV.

The brightest point was offset from the nucleus by 11\,000~km, dimming to 10\% of this value at a distance of 60\,000~km.

The \textsc{acis-S1 ccd} was not activated in this observation. The low quantum efficiency of the other activated \textsc{ccd}s below 0.5~keV makes them unsuitable as background references.

\noindent\textbf{Solar wind.} The proton velocity decreased during
observations from 600~\kms{} to 500~\kms. A flare on 20 November
2003, at N01W08 (FF\#525), was well-positioned to affect the
observations on 23 November (data from work in progress by Z.K.
Smith et al.). The comet most likely interacted with the
overexpanded, rarified plasma flow that followed the earlier hot
shocked and compressed flow behind the flare's shock.

\subsection{C/2001 Q4 (\textsc{neat})}
\noindent\textbf{X-rays.} A short observation of comet C/2001~Q4 was
made on May 12 2004, when the geocentric and heliocentric
distances were $\Delta = 0.362$~AU and $r_h = 0.964$~AU
respectively. With a heliographic latitude of 3 degrees, the comet
was almost in the ecliptic plane. From 3~pointings, the useful
exposure was 10\,328~seconds. The \textsc{acis-S3} chip detected 6540 X-rays in between 300 and
1000~eV. The \textsc{acis-S1} was used as a background signal.\\

\noindent\textbf{Solar wind.} There was no significant solar
activity during the observations (Z.K. Smith et al., ibid.). From
solar wind data, the comet interacted with a quiet, slow 352~\kms{}
wind.

\subsection{9P/2005 (Tempel~1)}
\noindent\textbf{X-rays.} The observation of comet 9P/2005~(Tempel
1) was designed to coincide with the {\em Deep Impact} mission
\cite{Lis07}. The allocated observation time of 291.6~ks was
split into 7~periods, starting on June 30th, July 4th
(encompassing the Deep Impact collision), July 5th, July 8th, July
10th, July 13th and July 24th. The brightest observing periods
were June 30th and July 8th. The focus here is on the June 30th
observation. On this date, $r_h = 1.507$~AU and $\Delta =
0.872$~AU.

The useful exposure was 50\,059 seconds, with a total of 7300
counts, 4000 from the June 30th flare alone, were detected in the
energy range of 300--1000~eV.

The brightest point for the June 30th observation was located 11\,000~km from the nucleus. The morphology appears to be more spherical than in other comet observations.\\

\noindent\textbf{Solar wind.} Observations were taken over a long
time span covering different solar wind environments. There was no
significant solar activity during the 30 June 2005 observations
(Z.K. Smith et al., ibid. \cite{Lis07}). From the \textsc{ace} data,
it can be seen that at June 30, the comet most likely interacted
with a quiet, slow solar wind.

\subsection{73P/2006 (Schwassmann--Wachmann 3B)}
\noindent\textbf{X-rays.} The close approach of comet 73P/2006
(Schwassmann--Wachmann~3B) in May 2005 ($\Delta = 0.106$~AU, $r_h =
0.965$~AU) provided an opportunity to examine cometary X-rays in
high spatial resolution. {\em Chandra} was one of several X-ray
missions to focus on one of the large fragments of the comet.
Between 300 and 1000~eV, 6285~counts were obtained in a
useful exposure of 20\,600~seconds.\\

\noindent\textbf{Solar wind.} There was a weak flare on 22 May
2006 (FF\#655, Z.K.~Smith, priv. comm.). A sequence of three high
speed coronal hole streams passed the comet in the period around
the observations and a corotating \textsc{cir} might have reached the comet
in association with the observations on 23 May, which is confirmed
by the mapped solar wind data.

\end{document}